\documentclass{emulateapj}

\slugcomment{{\sc Accepted to ApJ:} August 4, 2009}
\shorttitle{Compact Star Clusters in the M31 Disk}
\shortauthors{Vansevi\v{c}ius et al.}

\begin{document}

\title{Compact Star Clusters in the M31 Disk}

\author{V.~Vansevi\v{c}ius\altaffilmark{1,2}, K.~Kodaira\altaffilmark{3},
D.~Narbutis\altaffilmark{1,2,4}, R.~Stonkut\.{e}\altaffilmark{1}, \\
A.~Brid\v{z}ius\altaffilmark{1}, V.~Deveikis\altaffilmark{2}, and
D.~Semionov\altaffilmark{1}}

\altaffiltext{1}{Institute of Physics, Savanori\c{u} 231, Vilnius
LT-02300, Lithuania; donatas.narbutis@ff.vu.lt}
\altaffiltext{2}{Vilnius University Observatory, \v{C}iurlionio 29,
Vilnius LT-03100, Lithuania}
\altaffiltext{3}{The Graduate University for Advanced Studies
(SOKENDAI), Shonan Village, Hayama, Kanagawa 240-0193, Japan}
\altaffiltext{4}{Author to whom any correspondence should be addressed.}

\begin{abstract}
We have carried out a survey of compact star clusters (apparent size
$\lesssim$3\arcsec) in the southwest part of the M31 galaxy, based
on the high-resolution Suprime-Cam images ($17.5\arcmin \times
28.5\arcmin$), covering $\sim$15\% of the deprojected galaxy disk
area. The $UBVRI$ photometry of 285 cluster candidates ($V \lesssim
20.5$\,mag) was performed using frames of the Local Group Galaxies
Survey. The final sample, containing 238 {\it high probability} star
cluster candidates (typical half-light radius $r_{h} \sim 1.5$\,pc),
was selected by specifying a lower limit of $r_{h} \gtrsim 0.15\arcsec$
($\gtrsim 0.6\,{\rm pc}$). We derived cluster parameters based on
the photometric data and multiband images by employing simple stellar
population models. The clusters have a wide range of ages from
$\sim$5\,Myr (young objects associated with 24\,$\mu$m and/or
$H\alpha$ emission) to $\sim$10\,Gyr (globular cluster candidates),
and possess mass in a range of $3.0 \lesssim \log(m/m_{\sun}) \lesssim
4.3$ peaking at $m \sim 4000\,m_{\sun}$. Typical age of these
intermediate-mass clusters is in the range of $30\,{\rm Myr} \lesssim
t \lesssim 3$\,Gyr, with a prominent peak at $\sim$70\,Myr. These
findings suggest a rich intermediate-mass star cluster population
in M31, which appears to be scarce in the Milky Way galaxy.
\end{abstract}


\section{Introduction} \label{s:intro}
The Andromeda galaxy, M31, as the nearest spiral galaxy, has been
regarded to be the most suitable to provide supplementary data to
the Galactic data for understanding the galaxy structures, and to
further elucidate the evolution history of disk galaxies. Since a star
cluster normally represents a single population, many detailed studies
of star clusters in M31 were devoted to deriving information related
to the structure and evolution of star clusters and of hosting galaxy:
see the review in \citet{Kodaira2002} and the introduction of
\citet{Caldwell2009}, and the works cited therein, among others.

Recently, many observational works were devoted to study globular
clusters \citep[GCs; see, e.g.,][and references therein]{Barmby2002,
Barmby2007}. The study by \citet{Cohen2005}, based on high-resolution
Keck imaging, however, suggests that M31 cluster samples \citep[e.g.,][]
{Galleti2007} might be heavily contaminated by asterisms. Nevertheless,
the discussion on this issue by \citet{Caldwell2009} states that
appearance of low-mass clusters dominated by several bright stars
depends strongly on the photometric passband used, and emphasizes
the need of multiband observations to identify clusters correctly.

M31 disk cluster studies benefited from the high-resolution {\it Hubble
Space Telescope} ($HST$) images, which were used by \citet{Williams2001a,
Williams2001b} to identify young star cluster candidates. Recently
an extensive archival $BVI$ image survey of $HST$ WFPC2/ACS instrument
fields, which are scattering over a wide area of M31 in patches,
was conducted in search of new clusters \citep[see][and references
therein]{Krienke2007,Krienke2008}.

However, until the recent years, when high-resolution wide-field
images became available, much less efforts have been devoted for a
homogeneous wide-field photometric survey of objects, which have
properties in-between of globular and open clusters. Our survey,
using Suprime-Cam \citep{Miyazaki2002} mosaic images, completely
covers the area of $17.5\arcmin \times 28.5\arcmin$ in the southwestern
part of the M31 disk \citep[see][]{Kodaira2004}. A dozen of $HST$
frames are located in our field covering less than 20\%.

Our initial study of M31 compact clusters of an apparent size less
than $\sim$3\arcsec\ using Suprime-Cam revealed that some of the
previously suspected GC candidates should be classified as open
clusters (OCs), suggesting that there is a new kind of star cluster,
which is not well known in the Milky Way galaxy (MW), but whose
counterparts are probably present in Large Magellanic Cloud/Small
Magellanic Cloud \citep[LMC/SMC;][]{Kodaira2002,Kodaira2004,Kodaira2008}.
They are apparently more massive than typical OCs in the solar
neighborhood \citep{Piskunov2008}, but less massive than the typical
GCs in MW \citep{McLaughlin2005}. This suggests the presence of an
intermediate disk population of star clusters, a detailed study of
which may contribute to the understanding of disk evolution history.
For examples of the typical images of compact clusters, see the atlas
published by \citet{Kodaira2004}.

About 50 bright ($17.0 \lesssim V \lesssim 19.0$\,mag) members of
our M31 cluster sample \citep{Narbutis2006a} were studied in detail
concerning their radial surface brightness profiles using a tidal-cutoff
King model \citep {King1962} and a power-law EFF model \citep{Elson1987}.
Their distribution in the half-light radius versus the absolute $V$-band
magnitude diagram \citep{Sableviciute2007} shows that these compact
clusters are occupying the same domain as faint GCs \citep[e.g.,][]
{Barmby2007}. A comparison in the parameter domain of the King model
\citep{Sableviciute2006} confirms a visual impression that compact
clusters have similar radial profiles to GCs, except for minor
irregularities, which become visible in high-resolution images.
Comparison of the EFF model parameters of compact clusters and massive
clusters in LMC \citep{Mackey2003} and SMC \citep{Hill2006} shows
that the present sample of compact clusters may have the same
structural nature as massive clusters in the Magellanic Clouds
\citep{Sableviciute2006}.

These promising results motivated us to extend the survey of clusters
to a deeper magnitude of $V \lesssim 20.5$\,mag. The homogeneous
$UBVRI$ photometric data and multiband images for 285 selected cluster
candidates with absolute magnitudes down to $M_{V} \lesssim -4$\,mag
were analyzed in the catalog paper by \citet{Narbutis2008}. Here,
we present the results of simple stellar population (SSP) model
fitting using the multiband photometry data, giving mass and age
of these objects in the surveyed field of M31.

We have developed and applied a method to derive structural parameters
from the surface brightness distribution of clusters by properly
accounting for the contaminating stars, superposed on their smooth
profile. The derived half-light radii were studied in connection
with the SSP model fitting results. We have also modeled stochastic
effects of bright stars \citep{Deveikis2008} and briefly discuss
their influence on the derived star cluster parameters.

In the following, we describe observations and sample selection in
Section \ref{s:obs}, present structural and evolutionary star cluster
parameters in Section \ref{s:model_fit}, and analyze properties of
studied sample of M31 cluster candidates in Section \ref{s:results}.
Properties of a few particular objects are discussed in the Appendix.

\section{Observation} \label{s:obs}
\subsection{Data} \label{ss:data}
We have identified {\it high probability} star cluster candidates
(hereafter star clusters) by visually inspecting the Suprime-Cam
$V$-band mosaic image made from $5 \times 2$\,minute individual exposures
of the $17.5\arcmin \times 28.5\arcmin$ field, centered at $\alpha_{\rm
J2000}=0^{\rm h}40.9^{\rm m}$, $\delta_{\rm J2000}=+40\arcdeg45\arcmin$.
The characteristic full width at half maximum (FWHM) of stellar
images (point-spread function, PSF) is 0.7\arcsec. See \citet{Kodaira2004}
for a description of the Suprime-Cam observations and data reduction
details.

We have produced the homogenized $UBVRI$ photometric catalog of 285
selected star clusters ($V \lesssim 20.5$\,mag) used as {\it an
initial} sample in this study, based on photometry performed on the
Local Group Galaxies Survey (LGGS) mosaic images \citep{Massey2006}.
The variable PSFs of four overlapping LGGS field (F6--F9) mosaic
images were homogenized to the same ${\rm FWHM} = 1.5$\arcsec. The
mosaics were photometrically calibrated by using individual CCD
color equations and the stellar photometry catalog from \citet{Massey2006}
as a local photometric standard for zero point calibration. Different
aperture sizes and photometric background estimation areas were
selected for individual clusters, as they are given in the catalog
paper \citep{Narbutis2008}.

We note that a comparison of three published stellar photometry data
sets in our survey field \citep{Narbutis2006b} makes one to be
cautious when using tertiary standards as local photometric standards.
However, we used a carefully reduced and calibrated data set by
\citet{Massey2006}, which passed an internal consistency check,
making it the most accurately calibrated photometry catalog of the
M31 galaxy to date.

The color images of clusters were produced from the LGGS ($U, B, V,
I, H\alpha$), {\it GALEX} (FUV, NUV), 2MASS ($J, H, K_{\rm s}$),
{\it Spitzer} (24\,$\mu$m), and HI (21\,cm) images. They served as
references for object properties and environment study. Seventy-seven
star clusters from our catalog are located in the $HST$ archive
frames. For object selection, photometry, and multiband image details,
see \citet{Narbutis2008}.

We used Suprime-Cam $V$-band mosaic image (${\rm FWHM}_{\rm PSF}
\approx 0.7$\arcsec, image scale $0.2\arcsec/{\rm pixel}$) to derive
structural parameters of star clusters. The LGGS field's F7 $V$-band
mosaic image (${\rm FWHM}_{\rm PSF} \approx 0.8$\arcsec, image scale
of $0.27 \arcsec/{\rm pixel}$) was substituted for objects saturated
or having defects in the Suprime-Cam frame. Since the variability
of ${\rm FWHM}_{\rm PSF}$ was only $\sim$5\% across the mosaic (see
Section \ref{ss:str_pars} for details), we have applied the DAOPHOT
package \citep{Stetson1987} from the IRAF program system \citep{Tody1993}
to compute a single mosaic PSF. IRAF's program ``seepsf'' was used
to create PSFs suitable for our structural parameter fitting program.

The following M31 parameters are adopted in this study: distance
modulus of $m - M = 24.47$ \citep[785\,kpc, thus $1\arcsec \equiv
3.8$\,pc;][]{McConnachie2005}; center coordinates $\alpha_{\rm J2000}
=0^{\rm h}42^{\rm m}44.3^{\rm s}$, $\delta_{\rm J2000}=41\arcdeg
16\arcmin09\arcsec$ (NASA Extragalactic Database); major axis position
angle of 38\arcdeg\ \citep{Vaucouleurs1958}; disk inclination angle
to the line of sight of 75\arcdeg\ \citep{Gordon2006}.

\subsection{Sample Selection} \label{ss:sample}
The star clusters were selected from the {\it initial} sample of
285 objects by considering the following criteria: (1) the derived
half-light radius is of $r_{h} \geq 0.15\arcsec\,(\geq 0.6\,{\rm
pc}$)---this limit arises due to the resolution of ground-based
observations; (2) there are no nearby contaminants strongly influencing
the accuracy of the derived structural and evolutionary parameters.
Note however, that several genuine star clusters, judged from the
$HST$ images, did not meet these criteria.

A few dim, i.e., low central surface brightness, objects with an
unreasonably large derived half-light radius or composed of several
resolved stars were rejected because of low reliability of determined
parameters. Although P\'{E}GASE \citep{Fioc1997} SSP models incorporate
nebular emission lines for young age objects, we stress that structural
and evolutionary parameters of clusters showing up $H\alpha$ emissions
or superposed on inhomogeneous diffuse $H\alpha$ background should
be considered with caution. Note however, that objects with strong
$H\alpha$ emission were cataloged separately by \citet{Kodaira2004},
therefore, these objects were excluded from the analysis in this
study. We note, that foreground starburst galaxies possessing 24\,$\mu$m
emission, which have been revealed during spectroscopic study by
\citet{Caldwell2009}, might be present in a sample of young ($t
\lesssim 20$\,Myr) clusters (see Appendix for details on KW271 and
KW279).

This cleaning yielded a sample of 238 star clusters. Their ID numbers
and center coordinates from \citet{Narbutis2008} are supplemented
with the structural and evolutionary parameters described in the
following, and are presented in Table~1, the full form of which is
available in the electronic edition of the {\it Journal}.

\section{Model Fitting} \label{s:model_fit}
\subsection{Structural Parameters} \label{ss:str_pars}
The studied star clusters, projected on the crowded disk of M31,
are unresolved or semiresolved in the Suprime-Cam images. Background
and/or clusters' bright stars superposed on clusters' surface brightness
distribution strongly influence photometric \citep{Narbutis2007b}
and structural parameter accuracy. However, stars projecting nearby
to the cluster cannot be simply masked out because of flux superposition
in their extended wings. To avoid this, we have developed a method
to derive structural parameters of star clusters by including resolved
stars in the cluster model fitting procedure. The method is presented
in D. Semionov et al. (2009, in preparation); here we provide its brief
description and derived half-light radii of star clusters (Table~1).

The developed program tool models star clusters as a smooth analytical
surface brightness distributions. Observational effects are taken
into account by convolving cluster models with PSF. Individual resolved
stars are modeled with PSF. Free parameters, which describe a shape
and position of the cluster model, are fitted using a genetic algorithm,
providing the global solution. Fine tuning of this solution is performed
using Levenberg--Marquardt nonlinear least-square algorithm. A photometric
background value, which is kept constant during model fit, is estimated
within individually selected circular annulus centered on the object.
The model fitting radius is chosen seeking to enclose extended wings
of the surface brightness distribution. The residual images are
analyzed to ensure that there are no visible systematic deviations.

We have fitted 285 clusters from the {\it initial} sample with circular
King \citep{King1962} and EFF \citep{Elson1987} models. Since these
models have different sensitivity to the cluster's ``core'' and
``wings,'' comparison of the half-light radii derived by their means
provides the robustness of the model fit and $r_{h}$. Elliptical
models were not used in this study due to stochastic distribution
of bright cluster stars, which introduce fake ellipticity for
the low-mass clusters (see \citealt{Deveikis2008} for a discussion
of stochastic effects). A number of contaminating stars used for
the model fitting was: 0 in 40\%, 1 in 30\%, and 2 in 20\% cases.
The maximum number of fitted stars was 6.

Typical fitted values of structural parameters are: (1) the core
radius, $r_{c} \sim 0.75$\,pc, the concentration parameter, $c \sim
1.5$ (the King model); (2) the effective radius, $r_{e} \sim 0.75$\,pc,
and the power-law index, $n \sim 1.3$ (the EFF model). For $\sim$70\%
of clusters the derived $c$ and $n$ parameter values were confined
to the reasonable limits of $0.5 \leq c \leq 2.5$ \citep[see boundaries
for MW GCs in][]{Harris1996} and $1.05 \leq n \leq 3.5$, respectively.
For practical convenience, the remaining $\sim$30\% of clusters, having
determined parameters outside these limits, were re-fitted by setting
particular parameters to the corresponding boundary values. We note
however, that although this slightly affects derived $r_{c}$ or $r_{e}$,
it does not change the derived half-light radii nor introduces
systematic deviations in the residual images.

Finally, {\it intrinsic} half-light radii of objects, based on the
King, $r_{h}^{\rm K}$, and the EFF, $r_{h}^{\rm E}$, models were
derived. They are plotted in Figure~1(a) and show a good correlation.
The average half-light radii values, $r_{h} = (r_{h}^{\rm K} +
r_{h}^{\rm E})/2$, for 238 star clusters are provided in Table~1.

To estimate both the robustness of the model fitting and the uncertainty
of the derived half-light radius, $\sigma_{r_{h}}$, we have performed
two tests: (1) varied photometric background value, $\mu_{\rm sky}$,
which is kept constant during model fitting; and (2) decreased fitting
radius, $r_{\rm fit}$, from the adopted initial value. These tests
indicate that $r_{h}$ based on the King and the EFF models are in
a good agreement, and the overall accuracy of the $r_{h}$ value is
better than 20\%. The typical $\sigma_{r_{h}}$ is indicated by error-bars
in Figure~1(a). Based on these results, we applied a conservative lower
limit (which is mainly a subject to the resolution of ground-based
observations) of $r_{h} = 0.15\arcsec$\,($0.6\,{\rm pc}$) for the
sample selection. Note however, that some clusters below this limit
have been identified in the $HST$ images, which turned out to be of
a very compact nature or dominated by a central bright star.

To test the influence of PSF variability on the derived half-light
radii, we have divided survey field into 15 regions and constructed
individual PSFs for these regions. These PSF models were treated as
``virtual objects'' and analyzed as star clusters in attempt to
derive their structural parameters. The derived half-light radii of
eight ``virtual objects'' are displayed with asterisks in Figure~1(a)
and are confined in the range of $r_{h} < 0.4$\,pc. Note that only
eight of 15 ``virtual objects'' are visible, since others failed to
be fitted due to FWHM smaller than that of the adopted mosaic PSF.
The scatter of $r_{h} < 0.4$\,pc corresponds to the $r_{h}$ uncertainty
of 20\% for objects with $r_{h} \sim 2$\,pc. It is the upper limit
of uncertainty arising due to PSF variability across the survey field.
For some individual clusters, we have additionally derived structural
parameters by using local and mosaic PSFs, and found results being
in a good agreement.

\begin{figure}
\epsscale{1.1}
\plotone{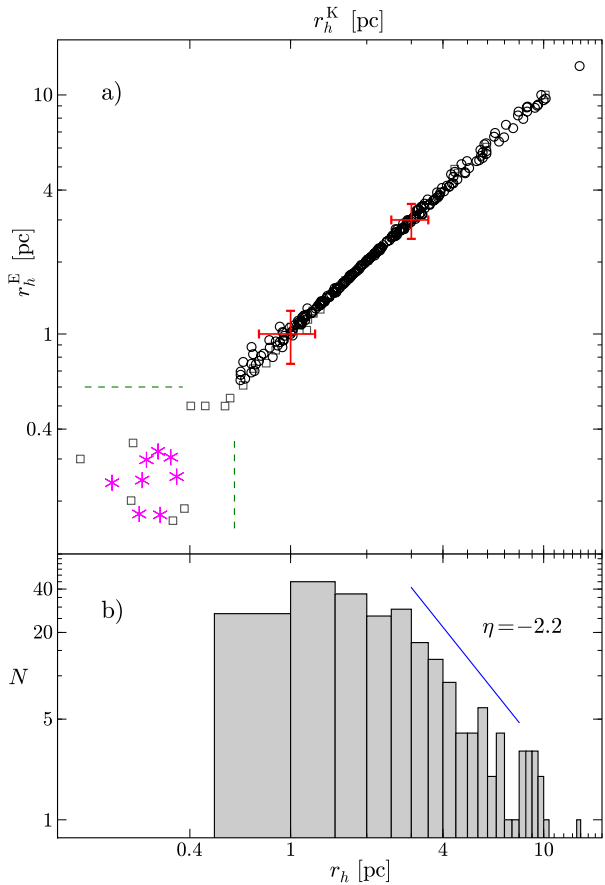}
\caption{Half-light radii of star cluster candidates. (a) Determined
half-light radii of 285 {\it initial} sample objects---based on the
EFF model, $r_{h}^{\rm E}$, and the King model, $r_{h}^{\rm K}$
(squares mark rejected cluster candidates; circles---selected 238
clusters; asterisks---PSFs constructed in the survey field and analyzed
as ``virtual objects;'' dashed lines---the applied half-light radii
selection limit $r_{h} \gtrsim 0.6$\,pc; error-bars---typical
$\sigma_{r_{h}}$). (b) Histogram of average half-light radii, $r_{h}
= (r_{h}^{\rm K} + r_{h}^{\rm E})/2$, of 238 clusters, overplotted
with a slope $\eta = -2.2$ line, defined as $N(r_{h}){\rm d}r_{h}
\propto r_{h}^{\eta}{\rm d}r_{h}$.}
\end{figure}

\begin{figure}
\epsscale{1.1}
\plotone{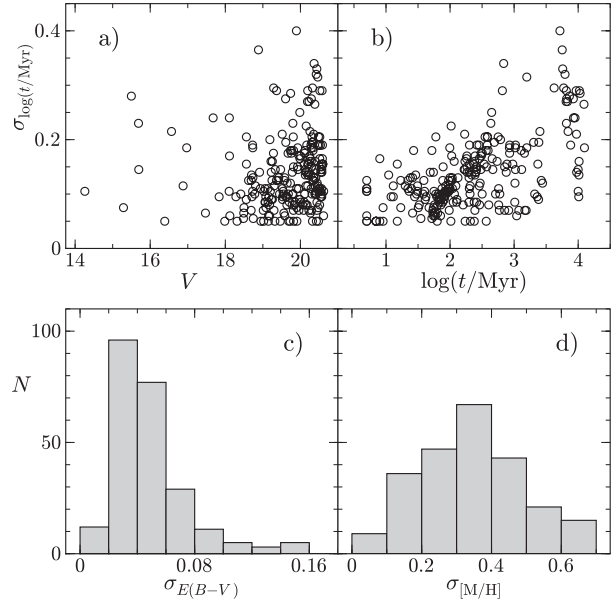}
\caption{Accuracy of the derived evolutionary parameters of 238 star
clusters. Panels (a) and (b) show the standard deviation of age,
$\sigma_{\log(t/{\rm Myr})}$, plotted vs. $V$-band magnitude and vs.
age, $\log(t/{\rm Myr})$, respectively. Panels (c) and (d) display
the histograms of standard deviations: color excess, $\sigma_{E(B-V)}$,
and metallicity, $\sigma_{\rm [M/H]}$, respectively.}
\end{figure}

In case of semiresolved star clusters, subtraction of contaminating
stars has a strong effect on the accuracy of derived half-light radii.
We have automatically included all stars within $r_{\rm fit}+{\rm
FWHM}_{\rm PSF}$ from the cluster's center and having magnitude
$\leq$$(m_{\rm cluster}+3)$\,mag, i.e., up to 3\,mag fainter than
the cluster itself, into the fitting procedure. Also some individual
strong contaminants were marked by hand. Conventional tests of varying
the background level and the fitting radius indicate that $\sim$20\%
accuracy of $r_{h}$ is achieved. Visual inspection of residual images
proved that clusters and contaminating stars are subtracted well.
We have estimated the goodness of model fitting by comparing the
standard deviation of the sky background within the cluster's area
after model subtraction and that of the nearby sky region. For
$\sim$90\% of cases, these values were equal, except for the cases
with larger residuals visible in the cluster's core, which, most
probably, arise due to semiresolved stars. They are the main source
of $r_{h}$ uncertainty for low-mass young star clusters \citep[see
Figure~3(c) in][]{Deveikis2008}.

We have compared $r_{h}$ values of 50 bright objects from our sample
with the ones derived by \citet{Sableviciute2007}. Despite different
analysis methods employed, we have found a good agreement between
these two studies. The maximum $r_{h}$ difference of $\sim$30\%
is attributed to the influence of contaminating stars not properly
accounted for in the previous study.

\subsection{Evolutionary Parameters} \label{ss:phot_pars}
The $UBVRI$ aperture CCD photometry data of 285 star clusters from
the {\it initial} sample \citep{Narbutis2008} were compared with
SSP models \citep[P\'{E}GASE code:][]{Fioc1997} assuming the standard
initial mass function \citep[IMF;][]{Kroupa2002} and various metallicities
to yield the {\it best fit} model evolutionary parameters, i.e.,
simple SSP model fitting was performed. The interstellar extinction
was accounted for assuming the standard extinction law \citep{Cardelli1989}.
The adopted parameter quantification technique and intrinsic precision
analysis, based on the $UBVRI$ integrated photometry, was presented
in detail by \citet{Narbutis2007a} and \citet{Bridzius2008}. This
technique, supplemented with a detailed investigation of multiband
images, introduced to reduce the age--metallicity--extinction degeneracy,
was used in the study.

The simple SSP model fitting results for 238 star clusters are provided
in Table~1. The determined evolutionary parameters are: absolute
$V$-band magnitude (corrected for aperture correction, see below
for details), $M_{V}$; age in Myr, $\log(t/{\rm Myr})$; mass in
solar-mass units, $\log(m/m_{\sun})$; metallicity, [M/H]; and color
excess, $E(B-V)$.

The standard deviation of age, $\sigma_{\log(t/{\rm Myr})}$, is
plotted versus $V$-band magnitude and age, $\log(t/{\rm Myr})$, in
Figures~2(a) \& (b), respectively. The standard deviation of age
increases typically from $\sim$0.1\,dex for young to $\sim$0.3\,dex
for old objects. We note that systematic differences of the derived
parameters could be expected if another SSP model bank would be used.
The histograms of the standard deviation of color excess, $\sigma_{E(B-V)}$,
and metallicity, $\sigma_{\rm [M/H]}$, are shown in Figures~2(c) \&
(d), respectively. The characteristic $\sigma_{E(B-V)}$ is $\sim$0.05,
reaching $\sim$0.15 for $t \sim 10$\,Gyr objects due to stronger
age--extinction degeneracy. The typical $\sigma_{\rm [M/H]}$ is
$\sim$0.3\,dex, reaching up to $\sim$0.7\,dex for $t \sim 200$\,Myr
objects due to stronger age--metallicity degeneracy (see \citealt{Narbutis2007a}
for parameter degeneracy maps).

We note that in this study numerous SSP model simplifications were
assumed: (1) a constant IMF; (2) a standard for MW and constant for
all cluster stars extinction law; (3) solar element ratio. Additionally,
in cases of unreliable photometric solutions for metallicity, we
restricted [M/H] determination within lower and upper boundaries as
a function of age taken from the chemical evolution model prediction.
Referring to models of MW and M31 by \citet{Renda2005}, we assumed
that metallicity evolution in the M31 survey field is not significantly
different from that of solar neighborhood, which has been modeled
by \citet{Schonrich2009}. Therefore, the [M/H] boundaries roughly
between $-1$ and $+0.5$ were considered, also taking into account
possibility that oldest objects could be globular clusters in M31
halo with [M/H] as low as $-2$.

\begin{figure}
\epsscale{1.1}
\plotone{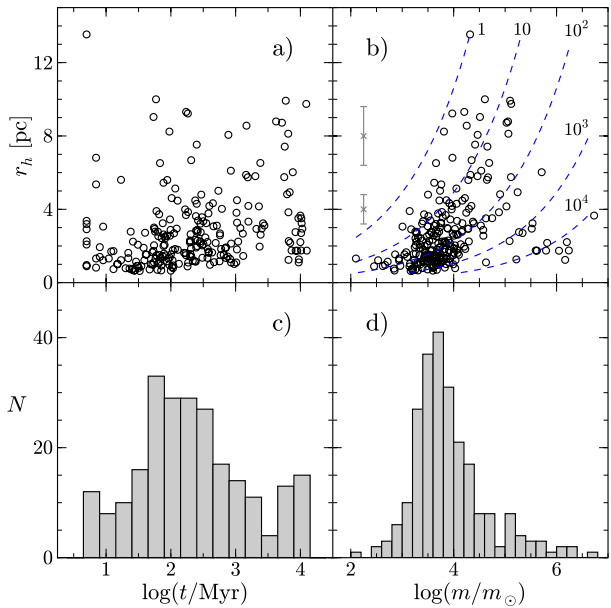}
\caption{Parameters of 238 star clusters. (a) Half-light radius,
$r_{h}$, plotted vs. age, $\log(t/{\rm Myr})$. (b) $r_{h}$ vs. mass,
$\log(m/m_{\sun})$, overplotted with lines of a constant half-mass
density, $\rho_{h} = 1$, $10$, $10^{2}$, $10^{3}$, and $10^{4}$
$m_{\sun}\,{\rm pc}^{-3}$, from left to right, respectively. (c)
and (d) Histograms of the derived cluster age, $\log(t/{\rm Myr})$,
and mass, $\log(m/m_{\sun})$, respectively. In panel (b) error-bars
indicate typical $\sigma_{r_{h}}$.}
\end{figure}

The mass of clusters in the solar-mass unit, $m/m_{\sun}$, was
calculated equating the mass-to-luminosity ratio of the {\it best
fit} SSP model of age, $t$, and metallicity, [M/H], to the absolute
$V$-band magnitude, $M_{V}$. The mass of clusters strongly depends
on the derived age, therefore, the primary sources of mass uncertainty
are the accuracy of age and extinction. Since cluster candidates
in the crowded field were measured through small apertures, we have
taken into account an individual aperture correction for each star
cluster by applying the following procedure. Artificial clusters
based on the best-fitted King model parameters were generated and
convolved with the homogenized PSF of ${\rm FWHM} = 1.5\arcsec$;
i.e., resolution of model cluster images was matched to the LGGS
frame resolution employed for cluster photometry. Individual aperture
used for a real cluster was applied to measure corresponding model
cluster, the difference between measured and total flux was assumed
as an aperture correction for a real cluster photometry. A typical
$V$-band aperture correction is of $\sim$0.4\,mag and translates
into a correction of mass by $\sim$0.15\,dex in $\log(m/m_{\sun})$
scale.

Analysis of star cluster population requires a proper assessment of
selection effects and resulting completeness. Stochastic modeling
of clusters \citep{Deveikis2008} unfolded a broad variety of their
possible appearances, indicating that visual selection of low-mass
clusters could be biased. However, for older clusters, i.e., after
the phase of supergiants ($t \gtrsim 100$\,Myr), this problem appears
to become less crucial.

\section{Results and Discussion} \label{s:results}
Here, we analyze the structural and evolutionary parameters of 238
star clusters in the southwestern field of the M31: half-light radius;
age; mass; color excess; metallicity; and spatial distribution. Prior
to presenting our results, we stress that important sources of systematic
bias and uncertainty of the derived parameters, in particular for
young low-mass star clusters, are stochastic effects \citep{Deveikis2008}.
Assuming stochastic nature of star formation to randomly populate
the stellar IMF, we modeled influence of stochastic effects on measurable
cluster characteristics, and found that structural and evolutionary
parameters could be biased---for some cluster age and mass intervals---in
a systematic way. Recently, \citet{Barker2008} gave a strong caution
that the standard SSP model analysis significantly underestimates
uncertainty in the derived cluster age. However, only standard SSP
model fitting was performed in this study; see a recent attempt to
solve this problem for star clusters of solar metallicity by
\citet{MaizApellaniz2009}.

\subsection{Half-Light Radius} \label{ss:rh}
The histogram of clusters' half-light radius, $r_{h}$, displayed
in Figure~1(b), spans the range from $\sim$0.6\,pc to $\sim$10\,pc
with one large object of $r_{h} \sim 14$\,pc (KW249), which is
described in the Appendix; the distribution peaks at $r_{h} \sim
1.5$\,pc. For comparison, the peak of the half-light radius distribution
of MW GCs from \citet[February 2003 rev.\footnote{See http://physwww.mcmaster.ca/$\sim$harris/mwgc.dat.}]
{Harris1996} catalog is at $\sim$2.5\,pc. Therefore, the majority
of clusters from our M31 sample are smaller (more {\it compact})
than typical MW GCs.

The turnover and a lack of small-size clusters close to the applied
lower limit of $r_{h} = 0.6$\,pc can be attributed to the visual
selection of the cluster sample. The exponential decrease in the
number of clusters in the larger half-light radii domain should
be real. Note however, tests performed with ``SimClust'' \citep{Deveikis2008}
show that only extended clusters of low luminosity ($M_{V} \sim -4.5$)
projected on star-forming regions could be missed by visual inspection.
The overplotted power-law slope $\eta = -2.2$ line, defined as $N(r_{h}){\rm
d}r_{h} \propto r_{h}^{\eta} {\rm d}r_{h}$, was found as the best
fit for M51 star clusters by \citet{Bastian2005}, and agrees well
with our star cluster size distribution in the range of $r_{h} = 3
- 8$\,pc. However, a considerably steeper slope was found in the
M51 cluster sample of \citet{Scheepmaker2007}, which is five times
more numerous than that of \citet{Bastian2005}.

The half-light radius of clusters is plotted versus age, $\log(t/{\rm
Myr})$, in Figure~3(a). No obvious half-light radius evolution with
the age of cluster population, as has been found in LMC/SMC by
\citet{Mackey2003}, can be judged from this diagram, since objects
with large $r_{h}$ are observed over a wide range of age. However,
the lack of small-size clusters at age $t \sim 3$\,Gyr can be attributed
to the selection effects discussed by \citet{Hunter2003} for the
LMC/SMC case, such that with increasing age only bright and, therefore,
massive extended clusters are selected in the magnitude limited sample.
At $t \sim 10$\,Gyr, halo GCs likely dominate in our sample and show
$r_{h}$ distribution similar to that of the MW GCs population with
$r_{h}$ as small as $\sim$1\,pc. They have high densities (see
Figure~3(b)) and likely have survived tidal disturbances by the disk.

The half-light radius versus mass, $\log(m/m_{\sun})$, distribution
is shown in Figure~3(b), overplotted with five lines of constant
half-mass density, $\rho_{h} = 3 m / 8 \pi r_{h}^{3}$, ranging from
1 to $10^{4}\,m_{\sun}\,{\rm pc}^{-3}$. The most numerous clusters
in the M31 southwestern field are {\it compact} clusters of small
size $r_{h} \sim 1.5$\,pc, intermediate mass $m \sim 4000\,m_{\sun}$,
and of young $t \sim 100$\,Myr age. Largest clusters seem to be
enveloped by the lines of $\rho_{h}$ from 1 to 100\,$m_{\sun}\,{\rm
pc}^{-3}$, lending an impression that massive clusters tend to be
bigger in size as far as clusters of relatively low specific density
of $\rho_{h} < 100$\,$m_{\sun}\,{\rm pc}^{-3}$ are concerned \citep{Hunter2003}.

Comparing the $r_{h}$ versus mass distribution of our sample with a
sample of M51 star clusters \citep{Scheepmaker2007}, we note a lack
of objects with $4 \lesssim \log(m/m_{\sun}) \lesssim 5$ and $r_{h}
\lesssim 5$\,pc in our sample. The most massive, $\log(m/m_{\sun})
\gtrsim 4.5$, clusters are ``branching'' in Figure~3(b) at small size,
$r_{h} \lesssim 3$\,pc, and high density, $\rho_{h} \sim
10^{4}\,m_{\sun}\,{\rm pc}^{-3}$, which is characteristic to old GCs.

The oldest clusters, which have highest densities, are most likely
halo GCs of small $r_{h}$. That the typical lifetime of $10^{4}\,m_{\sun}$
mass star cluster is estimated to be of $\sim$300\,Myr in the survey
field (see Section \ref{ss:form_disrp} for details) implies very low
probability for disk star clusters to reach an age of $\sim$10\,Gyr
without disruption. However, radial velocity measurements are necessary
to confirm whether these objects belong to halo.

\subsection{Age} \label{ss:age}
The distribution of clusters in the mass versus age diagram is shown
in Figure~4. The upper right domain of the diagram, $t \gtrsim 3$\,Gyr
and $4.0 \lesssim \log(m/m_{\sun}) \lesssim 7.0$, is occupied by
$\sim$30 classical GC candidates. The remaining $\sim$210 are most
likely disk clusters, which reside in the domain of $t \lesssim
3$\,Gyr and $\log(m/m_{\sun}) \lesssim 4.5$. The curve along the
lower envelope of cluster distribution is due to the selection limit
of $V \lesssim 20.5$\,mag, which was calculated with P\'{E}GASE SSP
models of ${\rm [M/H]} = -0.4$\,dex metallicity, assuming zero extinction.
Note, that star clusters are located above the selection limit due
to aperture corrections taken into account for mass calculation.
However, the upper envelope of the distribution should be free of
selection effects.

\begin{figure}
\epsscale{1.1}
\plotone{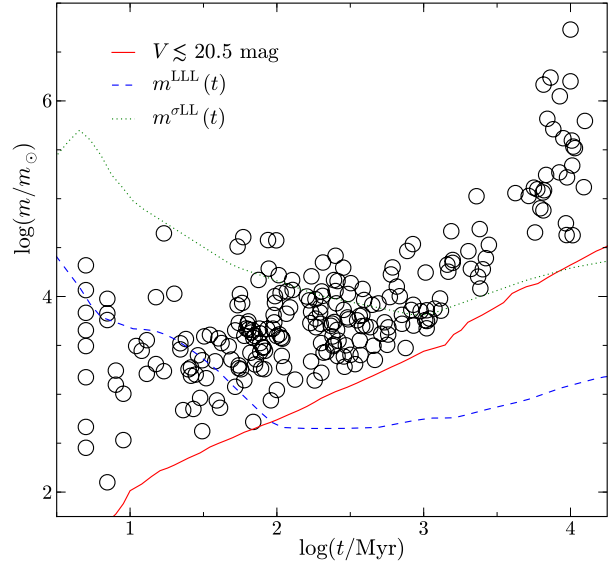}
\caption{Mass vs. age diagram of 238 star clusters, overplotted with
sample selection limit, $V \lesssim 20.5$\,mag, (solid line), the
Lowest Luminosity Limit, $m^{\rm LLL}(t)$, (dashed line), and 10\%
stochastic fluctuation of cluster $V$-band luminosity, $m^{\rm \sigma
LL}(t)$, (dotted line).}
\end{figure}

\citet{Cervino2004} have shown that SSP models cannot be used in a
straightforward deterministic way for small-mass cluster analysis,
when integrated luminosity of a model is lower than that of the
brightest star included in the used isochrone, i.e., ``Lowest Luminosity
Limit'' (LLL) requirement. They defined the smallest mass, $m^{\rm
LLL}$, associated with the LLL, and the mass, $m^{\rm \sigma LL}$,
corresponding to the 10\% stochastic fluctuation of the $V$-band
luminosity, $\sigma_{L_{V}} = 0.1 \times L_{V}$, which translates
into $\sim$0.04\,dex uncertainty of $\log(m/m_{\sun})$. We have
overplotted both parameters as a function of age in Figure~4, taken
from \citet{Cervino2004}, noting that $m^{\rm \sigma LL}(t) \sim
10 \times m^{\rm LLL}(t)$. Note however, that SSP models of solar
metallicity used by \citet{Cervino2004} do not significantly influence
a qualitative assessment of mass versus age distribution.

Judging from Figure~4, $m^{\rm LLL}$ is $\log(m/m_{\sun}) \sim 4$
for clusters younger than $\sim$10\,Myr. At the age of $\sim$50\,Myr,
$m^{\rm LLL}$ drops to $\log(m/m_{\sun}) \sim 3$, therefore, the
majority of clusters from our sample satisfy the LLL requirement.
At $t \sim 100$\,Myr, the $m^{\rm LLL}$ coincides with $V \lesssim
20.5$\,mag selection limit; older objects are well above LLL. Therefore,
clusters with $\log(m/m_{\sun}) \gtrsim 4.5$ should have the most
accurate evolutionary parameters.

The stochastic effects on the derived cluster mass can be estimated
by referring to the dotted line in Figure~4, based on the \citet{Cervino2004}
methods. The line displays 10\% uncertainty of cluster's luminosity
at a specific mass for a given age. Since we derived the mass from
the $V$-band luminosity, the mass as a function of age, having 10\%
uncertainty of mass, is depicted by the dotted line in Figure~4.
Formally, mass uncertainty increases to 100\% at the LLL indicated
by the dashed line. However, the uncertainty of mass is additionally
affected by the uncertainty of the derived age, which can be evaluated
from the slope of solid line in Figure~4, and also by the uncertainty
of extinction shown in Figure~2. Therefore, the LLL provides only
the lower limit for uncertainty arising due to stochastic effects,
which primarily bias age and extinction, and consequently---mass,
i.e., actual accuracy of evolutionary parameters is lower than it
is shown in Figure~2.

The histogram of cluster age, displayed in Figure~3(c), spans the range
from $t \sim 5$\,Myr to $\sim$12\,Gyr. Since evolutionary parameters
of the clusters with $H\alpha$ emission are less reliable due to
emission lines altering photometry and stochastic effects, we cannot
properly estimate parameters of the clusters younger than $\sim$20\,Myr.
Therefore, we did not specifically target our survey for such star
clusters.

The age distribution of clusters is displayed in Figure~5(a). We have
calculated the number of clusters per linear age interval, ${\rm
d}N/{\rm d}t$, by binning data in ${\rm d}\log(t/{\rm Myr}) = 0.2$,
0.3, and 0.5 age intervals, while also shifting them by a half-bin
width. These results were averaged and smoothed. The symbol size
indicates an uncertainty of the resulting distribution, which is
plotted versus age in a log scale. The age distribution is reasonably
well represented with two power-law lines of different slopes, and
is to be interpreted as a result of star cluster formation/disruption
processes (see Section \ref{ss:form_disrp} for discussion).

\begin{figure}
\epsscale{1.1}
\plotone{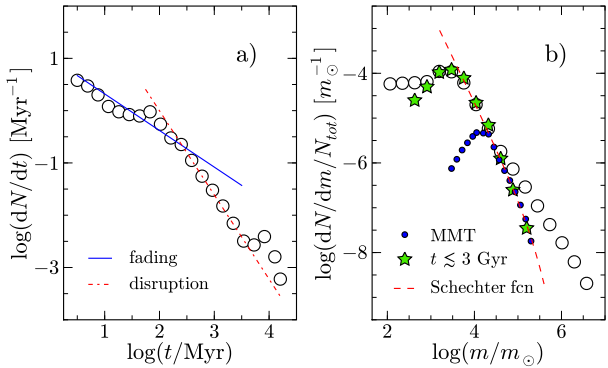}
\caption{Age and mass distributions of 238 star clusters, indicated
by open circles (size indicates the uncertainty of data binning) in
both panels. (a) Number of clusters per age interval, $\log({\rm
d}N/{\rm d}t)$, vs. age, $\log(t/{\rm Myr})$ (circles), overplotted
with power-law slopes expected for cluster population forming continuously
and: (1) fading evolutionary below the detection limit (solid line);
(2) instantaneously disrupting at some arbitrary time (dash-dotted
line). (b) Number of clusters per mass interval, normalized to the
total number of clusters in the sample, $\log({\rm d}N/{\rm d}m/N_{\rm
tot})$, vs. mass, $\log(m/m_{\sun})$ (circles), overplotted with
Schechter's function \citep[dashed line;][]{Gieles2009}, adopting
the turnover mass of $m^{*} = 2 \times 10^{5}\,m_{\sun}$. Stars
correspond to the intermediate age star clusters from our sample
in-between of $\sim$100\,Myr and $\sim$3\,Gyr; filled circles correspond
to the star clusters from MMT spectroscopic survey by \citet[][subsample
$V<19.0$\,mag, ages in-between of $\sim$100\,Myr and an
upper age limit of their sample $\sim$1\,Gyr]{Caldwell2009}.}
\end{figure}

The cluster number distribution ${\rm d}N/{\rm d}t$ (Figure~5(a))
in general follows the ``evolutionary fading'' line. However, after
subtracting the ``evolutionary fading'' number gradient, a gradual
increase in the number of clusters per log age interval from $t \sim
20$\,Myr to $t \sim 100$\,Myr is obvious. A peak of the cluster
number distribution is followed with a decrease in the number of
clusters, which is a consequence of an evolutionary cluster fading
effect, being always significant in the magnitude limited sample
(see Figure~4), and star cluster disruption. We note that analysis
of both Figures~3(c) \& 5(a) points to the peak of the cluster age
distribution at $t \sim 70$\,Myr, suggesting an enhanced cluster
formation episode at that epoch. A smaller secondary peak at $\sim$10\,Gyr
is due to old GC candidates.

\subsection{Mass} \label{ss:mass}
The histogram of cluster mass, displayed in Figure~3(d), spans the
range from $\sim$$10^{2}$ to $\sim$$5 \times 10^{6}\,m_{\sun}$ and
shows a prominent peak at $m \sim 4000\,m_{\sun}$. The decrease in
the number of clusters in the low-mass domain is apparently not
physical, but arises due to selection effects. The estimated completeness
of our cluster sample is of $\sim$50\% at $V \sim 20.5$\,mag
\citep{Narbutis2008}. Judging from these circumstances, the completeness
at the age of $\sim$100\,Myr and the mass of $\sim$4000\,$m_{\sun}$
(Figure~4) should be credible, even taking into account a stochastic
scattering of cluster luminosity at this age and mass. Therefore,
decline in the number of clusters per log mass interval in the
high-mass domain should be real.

The mass distribution of clusters is displayed in Figure~5(b). The
number of clusters per linear mass interval, normalized to the total
number of $N_{\rm tot} = 238$ clusters in our sample, ${\rm d}N/{\rm
d}m/N_{\rm tot}$, was calculated by binning data in ${\rm d}\log(m/m_{\sun})
= 0.2$, 0.3, and 0.5 mass intervals, and also shifting them by a
half-bin width. These results were averaged and smoothed. The symbol
size indicates the uncertainty of the resulting distribution. For
a more detailed comparison of intermediate age (from $\sim$100\,Myr
to $\sim$3\,Gyr) star clusters from our sample we have used the
selected subsample ($V<19.0$\,mag, age from $\sim$100\,Myr to an
upper age limit of their sample $\sim$1\,Gyr) of star clusters from
the M31 disk study by \citet{Caldwell2009}. The ${\rm d}N/{\rm d}m$
distribution presented in \citet{Caldwell2009} subsample was multiplied
by 0.15, taking into account that our survey field covers only $\sim$15\%
of the M31 disk. We note, that there are only eight objects in common
between our sample and that of \citet{Caldwell2009}.

Decline in the distribution of mass \citep{Caldwell2009} less than
$\log(m/m_{\sun}) \sim 4.3$ (see Figure~5(b)) arises due to a sample
incompleteness and resembles a shape of our intermediate age sample
distribution for mass less than $\log(m/m_{\sun}) \sim 3.5$. Although
we have only few star clusters in the high-mass range, the good match
with a ${\rm d}N/{\rm d}m$ slope \citep{Caldwell2009} and extension
down to the mass as low as $\log(m/m_{\sun}) \sim 3.7$ allows us
to constrain a shape of the star cluster mass function for age in-between
of $\sim$100\,Myr and $\sim$1\,Gyr. We overplotted Schechter's mass
function from \citet[Equation (1)]{Gieles2009}, adopting the turnover
mass of $m^{*} = 2 \times 10^{5}\,m_{\sun}$, which shows a good
agreement with both our intermediate age sample and \citet{Caldwell2009}
subsample data. We note that only vertical adjustment for the Schechter's
function was applied to match data at $m \sim 10^{4}\,m_{\sun}$.

Although we lack low-mass star clusters in our sample due to the
selection limit at $V \sim 20.5$\,mag ($M_{V} \sim -4.0$), the
detailed $HST$ study of star clusters in M31 by \citet{Krienke2007,
Krienke2008} has revealed a numerous population of star cluster
candidates as faint as $V \sim 23$\,mag ($M_{V} \sim -1.5$). They
are spatially well correlated with brighter counterparts of $t \sim
100$\,Myr from our sample, therefore, it is reasonable to assume
that they could be of a similar age. Thus, their mass should be
significantly smaller than $\sim$$4000\,m_{\sun}$, corresponding to
the mass range of typical MW OCs in the solar neighborhood \citep{Piskunov2008}.
Although the parameters of MW star clusters are derived basing on
individual stars, note that masses of MW clusters are a subject to
the accuracy of their derived tidal radii, as discussed by \citet{Piskunov2008}.

The high-mass domain of clusters from our sample is found to be
overlapping well with the sample of $\sim$140 young clusters, sharing
the M31 disk rotation, studied using MMT spectra and $HST$ images
by \citet{Caldwell2009}, who found a mass peak at $\log(m/m_{\sun})
\sim 4$, significantly higher than for our cluster sample, probably,
due to brighter cluster selection limit applied for spectroscopy.
The mass range of clusters from our sample younger than $\sim$3\,Gyr
falls in-between of the MW GCs \citep{McLaughlin2005} and the MW
OCs \citep{Piskunov2008}, indicating that we are mainly dealing
with the disk clusters from the intermediate mass range. The mass
distribution of clusters older than $\sim$3\,Gyr overlaps well with
the mass distribution of MW GCs \citep{McLaughlin2005}.

Comparing the mass versus age diagram of our cluster sample (Figure~4)
with those of LMC/SMC \citep{Hunter2003} and M51 \citep{Bastian2005},
we see a prominent lack of massive, $\log(m/m_{\sun}) \sim 5$, clusters
younger than $\sim$3\,Gyr in M31. Although our survey covers only
$\sim$15\% of the deprojected M31 disk, it contains a representative
part of its star-forming ring \citep{Gordon2006} in the vicinity
of NGC\,206. We also note that other studies of the M31 disk clusters,
which covered the whole galaxy, did not detect a significant number
of clusters more massive than $\log(m/m_{\sun}) \sim 4.5$ in this
age range either \citep[and references therein]{Caldwell2009}.

A few massive, $\log(m/m_{\sun}) \sim 5$, objects with an estimated
age of $t < 1$\,Gyr were reported by \citet{Caldwell2009}, who found
that most of their young clusters have a low concentration, typical
to the MW OCs. Their spatial distribution is subject to patchy field
selection of $HST$ high-resolution imaging over the M31 disk. Although
there are only eight objects in common in their and our samples,
judging from the overlapping mass range and the reasonable agreement
of the derived age, we may tentatively assume that both samples
belong to the same category of star clusters.

The general lack of star clusters of $\log(m/m_{\sun}) \sim 5$ mass
in the M31 disk could be due to a low-average star formation rate
in M31 \citep{Barmby2006}. However, an unfavorable inclination angle
of M31 could hide star clusters embedded in the M31 disk, mimicking
the situation in MW \citep{Clark2005}---there could be massive star
clusters hidden by dust clouds and yet to be detected in M31. This
hypothesis might be supported by the molecular cloud mass distribution
in M31 \citep{Nieten2006,Rosolowsky2007}, where star clusters with
a mass of up to $\log(m/m_{\sun}) \sim 5$ are expected to form, if
the typical star formation efficiency of $\sim$30\% in the cores of
molecular clouds is assumed.

\subsection{Extinction} \label{ss:ext}
A typical color excess of star clusters in our sample is $E(B-V)
\sim 0.25$. Clusters younger than $\sim$500\,Myr have a range of
$E(B-V) \lesssim 1.0$, which is consistent with a large extinction
of objects residing in the galaxy disk, e.g., four high extinction,
$E(B-V) \sim 1.1$, objects of $t \sim 100$\,Myr are projecting close
to the 24\,$\mu$m emission regions and prominent dust lanes. The
range of $E(B-V) \lesssim 0.4$ derived for clusters of intermediate
age, $500\,{\rm Myr} \lesssim t \lesssim 3$\,Gyr, might be related
to the evolutionary fading of clusters and the observed narrow mass
interval in this age range (see Figure~4), therefore, objects possessing
color excess higher than $\sim$0.4 are missing from our sample. We
note that $E(B-V)$ versus projected distance from the M31 center does
not show any tight correlation. Moreover, $E(B-V)$ of an individual
cluster seems to be more strongly dependent on the local interstellar
matter environment than on the radial distance from the M31 center.

The GC candidates of $t \gtrsim 3$\,Gyr have color excess over the
range of $E(B-V) \lesssim 0.6$, consistent with their presence in
the M31 halo on both sides of the disk. There are numerous objects
with $E(B-V) \sim 0.04$, which is in a good agreement with MW color
excess in the direction of the survey field in M31, estimated from
the MW extinction maps \citep{Schlegel1998}.

Note however, that for eight objects in common with \citet{Caldwell2009}
sample, which have $E(B-V) \sim 0.25$ derived in their study, we
determine color excess from 0.05 to 0.65. Although \citet{Caldwell2009}
determined extinction by examining the shape of spectra continuum,
for those cases when late-type stars dominate they used a mean color
excess for young clusters of $E(B-V) \sim 0.28$, which is in agreement
with the typical value of $E(B-V) \sim 0.25$ derived in our study
for presumably the same population of star clusters.

\subsection{Metallicity} \label{ss:metal}
The metallicity, [M/H], of star clusters, plotted versus age and
projected distance from the M31 center, $R_{\rm M31}$, in Figures~6(a)
\& (b), respectively, spans a range from $-2.0$ to $+0.5$\,dex. We
divided clusters into two age groups of $t \lesssim 3$\,Gyr and $t
\gtrsim 3$\,Gyr, indicated by a different symbol size in Figure~6(b).
The apparent narrowing of the metallicity scatter at an age of $\sim$200\,Myr
is attributed to the SSP model fitting artifact due to a strong
age--metallicity degeneracy in this age domain \citep{Narbutis2007a}.
Although the accuracy of photometrically derived metallicity is low
(see Figure~2(d)), the evolutionary trend with time is apparent.

\begin{figure}
\epsscale{1.1}
\plotone{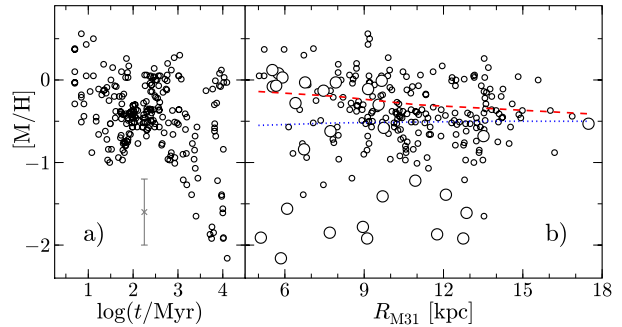}
\caption{Metallicity, [M/H], of 238 star clusters plotted vs. age,
$\log(t/{\rm Myr})$, and projected distance from the M31 center,
$R_{\rm M31}$, in panels (a) and (b), respectively. In panel (a)
error-bar indicates typical uncertainty of metallicity taken from
Figure~2(d). In panel (b) clusters are divided into two age groups
of $t \lesssim 3$\,Gyr (small circles) and $t \gtrsim 3$\,Gyr (large
circles), and overplotted with the average metallicity trends for
disk red giant stars from \citet[][dashed line]{Worthey2005} and
\citet[][dotted line]{Bellazzini2003}.}
\end{figure}

A rather constant average metallicity of ${\rm [M/H]} \sim -0.4$\,dex
is observed for clusters younger than $\sim$1\,Gyr, however, there
is a slight tendency of the average [M/H] to decrease with increasing
galactocentric distance. The radial profile of cluster metallicity
matches well those of disk red giant stars from \citet{Worthey2005}
and \citet{Bellazzini2003} shown by lines in Figure~6(b).

There are 20 clusters in our sample, which have [Fe/H] estimates
from the William Herschel Telescope spectra by \citet{Perrett2002}.
By accounting for $\alpha$-element overabundance typical to MW GCs,
we see a reasonable agreement between photometric and spectroscopic
metallicity estimates of old, $t \sim 10$\,Gyr, clusters. Two of
those metal-rich massive clusters (KW221 and KW225), presumably similar
to GCs, have metallicity of ${\rm [M/H]} \sim -0.8$ and 0.0\,dex,
and are found in our survey field area closest to the M31 bulge,
located at the projected distance of $R_{\rm M31} \sim 5$\,kpc from
the galaxy's center. Note however, that for several objects of $\sim$100\,Myr
age, the metallicity estimated by \citet{Perrett2002} is of ${\rm [Fe/H]}
\sim -2.2$\,dex, i.e., significantly lower than a photometrically
estimated ${\rm [M/H]} \sim -0.4$\,dex.

The elaborated M31 evolution models, such as ``M31 model-b'' by
\citet{Renda2005}, may well reproduce the metallicity gradient and
the metallicity change in time, as they are presented in Figure~6.
Detailed studies of the age-dependent profiles of the disk cluster
population would provide further constraints on the M31 disk
evolution models.

\subsection{Spatial Distribution} \label{ss:spatial}
Positions of star clusters are indicated on the {\it Spitzer}
(24\,$\mu$m) image of the M31 southwestern field in Figure~7. Clusters
are divided into three age groups: {\it young} ($t \lesssim 150$\,Myr);
{\it intermediate} ($150\,{\rm Myr} \lesssim t \lesssim 3$\,Gyr);
and {\it old} ($t \gtrsim 3$\,Gyr), shown by different symbols. The
distribution of {\it young} clusters resembles that of warm dust,
but is shifted outward to larger $R_{\rm M31}$ by $\sim$1.5\,kpc
along the galaxy's major axis. The {\it intermediate} age clusters
are smoothly distributed over the whole field. Old age clusters (GC
candidates) scatter over the area, but show a higher concentration
toward the M31 bulge.

\begin{figure}
\epsscale{1.1}
\plotone{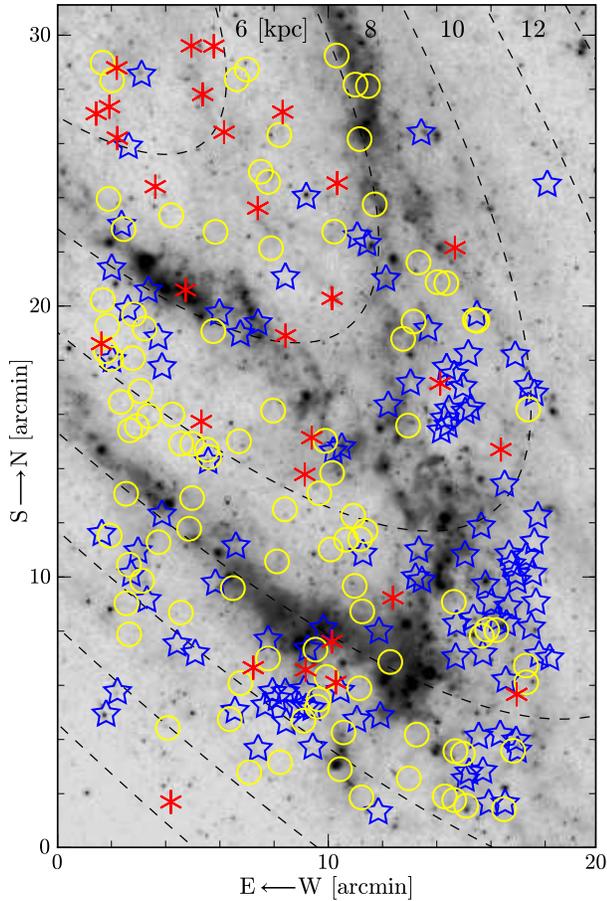}
\caption{Distribution of 238 star clusters in the M31 SW field,
overlaid on {\it Spitzer} (24\,$\mu$m) image. Clusters are divided
into three age groups: $t \lesssim 150$\,Myr (stars), $150\,{\rm
Myr} \lesssim t \lesssim 3$\,Gyr (circles), and $t \gtrsim 3$\,Gyr
(asterisks). Elliptical ring segments, indicating projected distance
from the M31 center, $R_{\rm M31} = 6 - 18$\,kpc, are marked with
dashed lines. North is up, east is left.}
\end{figure}

Star clusters with ultraviolet ({\it GALEX}) or 24\,$\mu$m ({\it
Spitzer}) emission are mainly distributed along the sites of active
star formation. At the bottom of the survey field, {\it young} and
{\it intermediate} age clusters appear to be well mixed. Around the
position of [6\arcmin,16\arcmin], where a lack of warm dust is
attributed to the ``split'' of the star forming ring, {\it intermediate}
age clusters dominate. However, around [17\arcmin,11\arcmin] mainly
{\it young} clusters are observed. The spatial density of {\it young}
and {\it intermediate} age clusters decreases toward the bulge,
which is consistent with a gas distribution in the M31 disk \citep{Nieten2006}.

To explain the M31 spiral structure and features of the prominent
star-forming ring at $R_{\rm M31} \sim 10$\,kpc, two scenarios of
the M32 galaxy having passed through the M31 disk were recently
proposed: (1) passage trough the disk occurred $\sim$20\,Myr ago
\citep{Gordon2006} and created a ``split'' of the ring around position
[6\arcmin,16\arcmin] (see Figure~7; NGC\,206 is located at the ring's
``branching'' point [15\arcmin,17\arcmin]); (2) passage trough the
M31 center occurred $\sim$210\,Myr ago \citep{Block2006}. The star
cluster population data could provide a reference to disentangle
the more preferable scenario.

True locations of objects in the M31 disk are subject to a strong
projection effect due to an unfavorable inclination angle of $\sim$75\arcdeg.
The projection of the galaxy and its rotation are such that the southwestern
part of the disk approaches the observer \citep{Henderson1979}; the
corotation radius lies beyond $R_{\rm M31} \sim 20$\,kpc \citep{Efremov1980,
Ivanov1985}; and the rotation period at $R_{\rm M31} \sim 10$\,kpc
distance is of $\sim$250\,Myr. Assuming that all clusters in our
sample are residing in the M31 disk, we made a deprojection of cluster
locations to face-on view, and referring to the disk rotation curve
\citep{Carignan2006}, computed object positions as a function of
look-back time.

Judging by eye, the clumps of {\it young} clusters, which are located
at [9\arcmin,5\arcmin] and [17\arcmin,10\arcmin], formed two compact
spatial configurations $\sim$$80 \pm 20$\,Myr ago. This age coincides
well with the cluster age peak (see Section \ref{ss:age}) and suggests
a possible evolutionary connection. There was no other compact spatial
cluster configuration noted in the look-back time range from $\sim$20\,Myr
to $\sim$150\,Myr, which is valid for this kind of study based on
our cluster sample. The range of the age distribution peak, estimated
from the SSP model fitting, is of $\sim$$50 - 120$\,Myr. However,
there is an age--metallicity degeneracy observed at the age of $\sim$100\,Myr,
which could slightly bias the age determination. Therefore, we note
this interesting coincidence of cluster age peak and compact spatial
cluster configurations. To discuss this finding more reliably, a
global kinematic study of the M31 disk cluster population and a
detailed dynamic model of the M31 and M32 encounter are needed. Note
however, that \citet{Caldwell2009} did not find evidence for a peak
in star formation between 20\,Myr and 1\,Gyr.

\subsection{Cluster Formation and Disruption} \label{ss:form_disrp}
If star clusters are being formed continuously at a constant rate
with a power-law cluster initial mass function, their number increases
with time in the mass versus age diagram as described by \citet{Boutloukos2003}.
Consequently, the mass of the most massive cluster in the older log
age bins rises due to a statistical effect of sample size. However,
in our sample, we see a lack of massive $\log(m/m_{\sun}) \gtrsim 4$
clusters older than $\sim$100\,Myr (Figure~4), probably, indicating
a non-constant cluster formation history. Therefore, using the observed
mass and age distributions of star clusters we attempt to derive
their formation and disruption rates in the studied field by adopting
the technique developed by \citet{Boutloukos2003}.

For a simple estimate, first we selected clusters more massive than
$\log(m/m_{\sun}) \sim 3.5$. There are $N_{1} \sim 30$ and $N_{2}
\sim 70$ objects in the age ranges $\Delta t_{1} \equiv 30 \lesssim
t \lesssim 100$\,Myr and $\Delta t_{2} \equiv 100\,{\rm Myr} \lesssim
t \lesssim 1$\,Gyr, respectively. If a cluster formation rate (CFR)
is constant and there is no cluster disruption, then a ratio of $N_{2}
/ N_{1} = \Delta t_{2} / \Delta t_{1} \approx 13$ is expected. The
actual ratio is $N_{2}/N_{1} \sim 2$, indicating that CFR is not
constant and/or significant cluster disruption took place during
the last Gyr in the M31 disk.

The number of clusters per linear age interval, ${\rm d}N/{\rm d}t$,
plotted versus age in Figure~5(a) can be described by two power-law
slopes, intersecting at $t \sim 300$\,Myr. The slope values were
taken from the fading/disruption models by \citet[Equation (14):
$\alpha=2.0$, $\gamma=0.62$, $\zeta=0.69$]{Boutloukos2003} and only
vertical adjustment to the observed distribution was applied. In
the age domain of $t \lesssim 300$\,Myr the slope line, expected
for a cluster population forming continuously and fading in the course
of evolution below the detection limit, provides a rather good match
to the observed data. Around $t \sim 70$\,Myr, there is a sudden
increase in the number of clusters by a factor of $\sim$2, which
could be attributed to the increased CFR at that epoch. In the age
domain older than $\sim$300\,Myr, the observed distribution can be
well described by the slope expected for a cluster disruption at
some arbitrary time. An increase in the cluster number at $t \sim
10$\,Gyr is attributed to the GC candidates, prominent in our sample.

The number of clusters per linear mass interval, ${\rm d}N/{\rm d}m$,
is displayed versus mass in Figure~5(b). In the mass range of $\log(m/m_{\sun})
\lesssim 3.3$ the incompleteness of our cluster sample is obvious.
Due to incompleteness, the mass distribution of our cluster sample
cannot be used to constrain the typical lifetime of star clusters
since the fading line cannot be adjusted properly. Note that the
distribution of clusters in between $\sim$100\,Myr and $\sim$3\,Gyr
and in the mass range of $\log(m/m_{\sun}) \gtrsim 4.5$ is steeper
than that of the whole sample, which includes the old GC candidates.

Using the \citet{Boutloukos2003} technique and relying on the power-law
slope line intersection point in Figure~5(a), we estimate a conservative
value of typical lifetime of $\log(m/m_{\sun}) = 4$ mass cluster to
be of $t_{4}^{\rm dis} \sim 300$\,Myr in the M31 survey field. For
comparison, the following values in other galaxies were derived by
\citet{Lamers2005}: M51 central region---$\sim$70\,Myr; M33---$\sim$630\,Myr;
MW solar neighborhood---$\sim$560\,Myr; and SMC---$\sim$8\,Gyr.

The cluster disruption analysis gives the disruption time of $\sim$500\,Myr
if only cluster age distribution of our sample is used. If the mass
distribution is taken into account (although it is difficult to define
intersection point of fading and disruption lines in Figure~5(b)),
the cluster disruption time can be estimated to be as small as $\sim$100\,Myr
since it is sensitive to the ``intersection'' mass. Therefore, we
choose $t_{4}^{\rm dis} \sim 300$\,Myr as a conservative value. We
note that a similar inconsistency between age and mass distribution
analysis was noticed recently for LMC star clusters by \citet{Parmentier2008}.
However, they found an opposite effect for the LMC sample---the
cluster mass distribution analysis tends to yield higher cluster
disruption timescale than the age distribution analysis. Such discrepancy
might be related to the assumptions applied in the \citet{Boutloukos2003}
analysis used in our study: (1) a constant cluster formation rate;
(2) a power-law cluster initial mass function, which should be of
the Schechter's function type as shown in Section \ref{ss:mass}.

\citet{Lamers2005} found anticorrelation between the lifetime,
$t_{4}^{\rm dis}$, of $\log(m/m_{\sun}) = 4$ mass cluster and the
ambient density of its environment, $\rho_{\rm amb}$, in the galaxies.
Following the reasoning by \citet{Lamers2005}, we attempt to estimate
$\rho_{\rm amb}$ in the M31 disk. We took the structural parameters
of the M31 disk from the ``best-fit model'' of \citet{Geehan2006}:
a disk scale-length, $R_{d} = 5.4$\,kpc; a central surface density,
$\Sigma_{0} = 4.6 \times 10^{8}\,m_{\sun}\,{\rm kpc}^{-2}$. Therefore,
the average disk surface density at $R_{\rm M31} \sim 10$\,kpc, i.e.,
at the center of our survey field $\Sigma = \Sigma_{0}\exp(-R_{\rm
M31}/R_{d}) \sim 70\,m_{\sun}\,{\rm pc}^{-2}$.

\citet{vanderKruit2002} has shown that for spiral galaxies the vertical
scale height is of $h_{z} \sim 0.15\,R_{d}$; this implies $h_{z}^{\rm
stars} \sim 800$\,pc. \citet{Braun1991} found $h_{z}^{\rm gas} \sim
350$\,pc for a gas distribution at a galactocentric distance of $R_{\rm
M31} = 10$\,kpc. The scale height for dust distribution is assumed
to be of $h_{z}^{\rm dust} \sim 100$ and $\sim$150\,pc by \citet{Hatano1997}
and \citet{Semionov2003}, respectively. Depending on the adopted
$h_{z}$, the estimated ambient density in our survey field, $\rho_{\rm
amb} \sim 0.5\, \Sigma\, h_{z}^{-1}$, is in the range of $\rho_{\rm
amb} \sim 0.05 \ldots 0.35\,m_{\sun}\,{\rm pc}^{-3}$. This density
is much lower than the typical half-mass density $\rho_{h} \sim
10^{2}\,m_{\sun}\,{\rm pc}^{-3}$ of star clusters, shown in Figure~3(b).

Referring to Figure~4 in \citet{Lamers2005}, we see that values of
$t_{4}^{\rm dis} \sim 300$\,Myr and $\rho_{\rm amb} \sim 0.05 \ldots
0.35\,m_{\sun}\,{\rm pc}^{-3}$ derived here, place the studied field
of the M31 disk in the position between MW and M33 and M51, however,
slightly below the predicted theoretical $t_{4}^{\rm dis} = t_{4}^{\rm
dis}(\rho_{\rm amb})$ line. \citet{Gieles2006} have modeled encounters
of star clusters with molecular clouds to explain short disruption
time for star clusters in the M51 central region. Our survey field
in M31 is centered on the star-forming ring possessing high-density
gas \citep{Nieten2006}, therefore, enhanced star cluster disruption
in respect to MW and M33 could apparently take place. Although
\citet[and references therein]{Rosolowsky2007} discuss similarity of
the molecular cloud properties in M31 and MW, enhanced star cluster
disruption should apparently take place in the M31 molecular gas
``ring'' in respect to that of the solar neighborhood. The whole
field survey of the M31 disk clusters and more detailed models of
cluster and cloud kinematics, which are needed to estimate encounter
cross sections in the disk, should be considered in future.

Finally, we attempt to estimate the CFR in the M31 disk. We select
$N = 51$ clusters with age $\Delta t \equiv 30\,{\rm Myr} \lesssim
t \lesssim 130$\,Myr and mass $3.3 \lesssim \log(m/m_{\sun}) \lesssim
4.5$ (see Figure~4). This subsample should have accurately estimated
parameters and be relatively free of selection and cluster disruption
effects. The total stellar mass of these clusters is $\Delta m \sim
3 \times 10^{5}\,m_{\sun}$. Assuming the constant power-law cluster
mass function \citep[index $-2.0$;][]{Gieles2009} down to $\log(m/m_{\sun})
= 2$, mass correction factor of $\sim$2.5 was deduced. Therefore,
the approximate rate of star formation in the clusters was ${\rm
CFR}_{\rm field} = \Delta m/ \Delta t \sim 0.008\,m_{\sun}\,{\rm
yr}^{-1}$. The deprojected area of survey field (Figure~7) is of
$\sim$70\,${\rm kpc}^{2}$, making $\sim$15\% of the galaxy disk.
From this, we estimate the lower limit of the average CFR at the
epoch of $\sim$100\,Myr to be of ${\rm CFR}_{\rm M31} \sim 0.05\,m_{\sun}\,{\rm
yr}^{-1}$ over the M31 disk. This can be compared to the present-day
star formation rate of 0.4\,$m_{\sun}\,{\rm yr}^{-1}$ \citep{Barmby2006},
indicating that $\gtrsim$10\% of formed stars could remain ``locked''
in star clusters during $\sim$100\,Myr.

For comparison, \citet{Williams2003} found the mean star formation
rate for the disk $\sim$1\,$m_{\sun}\,{\rm yr}^{-1}$ over the last
60\,Myr. Based on stellar population analysis, he suggests that the
lowest star formation rate occurred $\sim$25\,Myr ago being $\sim$2
times lower compared with the epoch from $\sim$50\,Myr to $\sim$250\,Myr.
Note, a good coincidence of star formation history in M31 deduced
by \citet{Williams2003} and one inferred from our cluster population
data (Figure~5(a)), which suggests an episode of a double increase
in cluster formation $\sim$70\,Myr ago. Also, \citet{Kodaira1999}
have found that the star formation rate in the field around OB
association A24 has decreased by $\sim$2.5 times from the epoch at
$\sim$1\,Gyr to the present day.

Although clusters in the intermediate mass range ($3.5 \lesssim
\log(m/m_{\sun}) \lesssim 4.5$) are not well known in the MW, the
compact clusters in M31 might represent a new class of star clusters
of intermediate age $30\,{\rm Myr} \lesssim t \lesssim 3$\,Gyr,
linking the age of GC formation and the present. It could be suspected
that this class of disk-population clusters is not observed in the
MW because of high extinction through the Galactic plane. This might
be supported by the \citet{Clark2005} discovery that a Galactic
star cluster Westerlund\,1 is a {\it Super Star Cluster} with a mass
of up to $\log(m/m_{\sun}) \sim 5$, age of $\sim$3\,Myr, and radius
of 0.6\,pc, which suffers strong interstellar extinction of $A_{V}
\sim 11.5$\,mag being at the distance of $\sim$5\,kpc. Therefore,
a more numerous massive compact star cluster population also could
be hidden by dust clouds in the M31 disk.

\section{Summary} \label{s:summary}
The sample of compact star clusters in the present homogeneous
photometric survey of the $17.5\arcmin \times 28.5\arcmin$ field in
the southwestern part of the M31 disk is apparently overlapping with
the clusters detected in the patchy $HST$ fields, in terms of their
mass $3.0 \lesssim \log(m/m_{\sun}) \lesssim 4.5$, which is in-between
of the mass of typical OCs and classical GCs, well known in the MW.
The structural properties of the compact clusters are similar to
those of GCs, except for minor irregularities, and they resemble
those of massive clusters in the LMC/SMC, including blue clusters
of globular appearance. The metallicity of sample clusters is rather
constant over a wide range of galactocentric distances, with a typical
value scattering around ${\rm [M/H]} \sim -0.4$\,dex.

The global follow-up survey of the whole M31 disk cluster population,
based on photometry and size measurement, is of great interest and
importance to comprehend the evolution of galaxy disks. Supplemented
with spectroscopic and higher image resolution studies, as well as
modeling of cluster stochastic effects, it would provide a guiding
panoramic view and better constraints for the M31 and M32 interaction
models.

\acknowledgments
We are grateful to the anonymous referee for constructive suggestions,
which helped to improve the paper considerably. This work was financially
supported in part by a Grant of the Lithuanian State Science and
Studies Foundation. The star cluster survey is based on the Suprime-Cam
images, collected at the Subaru Telescope, which is operated by the
National Astronomical Observatory of Japan. The research is based
in part on archival data obtained with the {\it Spitzer Space Telescope},
and has made use of the following: the NASA/IPAC Extragalactic Database
(NED) and the NASA/IPAC Infrared Science Archive, which are operated
by the Jet Propulsion Laboratory, California Institute of Technology,
under contract with the National Aeronautics and Space Administration;
the SAOImage DS9, developed by Smithsonian Astrophysical Observatory;
the USNOFS Image and Catalog Archive operated by the United States
Naval Observatory, Flagstaff Station. The data presented in this
paper were partly obtained from the Multimission Archive at the Space
Telescope Science Institute.

\section*{Appendix}
\section*{Objects for Detailed Study} \label{appendix:specific}
Here, we present several remarkable objects, which could serve as
targets for future spectroscopy and high-resolution imaging. We
briefly discuss their properties referring to the derived parameters,
multiband and $HST$ images. We provide cross-referencing with The
Revised Bologna Catalog of M31 Globular Clusters and candidates
compiled by \citet{Galleti2007} in brackets when available.

KW044 {\it (B325)} is located at the position of [16\arcmin,2\arcmin]
in Figure~7 and is resolved in Suprime-Cam images. The structural
model fit resulted in a large half-light radius, $r_{h} \sim 11$\,pc,
and an extremely small ratio of King model parameters, $r_{t}/r_{c}
\sim 3$. Its age of $\sim$60\,Myr, mass of $\log(m/m_{\sun}) \sim
4.6$, and radial velocity of $-$560\,km s$^{-1}$, compatible with
gas disk velocity at this position, imply that KW044 is a representative
of the young disk cluster population. However, another solution for
evolutionary parameters (a low extinction case) also exists---the
cluster could be an old object of low concentration and low metallicity,
resembling the extended ones in M31 \citep{Mackey2006} and M33
\citep{Stonkute2008}, and suggesting its location in the M31 halo.
The age of $\sim$630\,Myr is derived by \citet{Caldwell2009}.

\begin{figure}
\epsscale{1.1}
\plotone{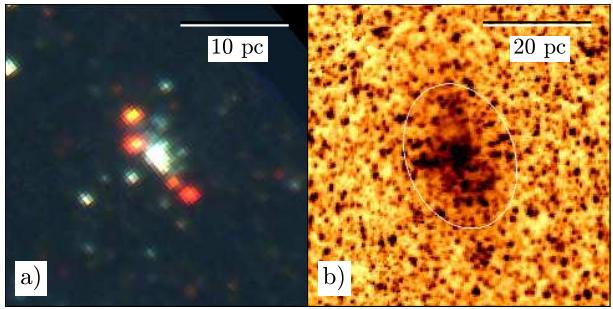}
\caption{$HST$ images. Panels show: (a) KW141---an example of stochastic
appearance of red bright stars ({\it color coding of $HST$ filters
in the electronic edition: red---F814W, green---F555W, and blue---F439W});
(b) KW249---an object with the largest half-light radius, $r_{h}
\sim 14$\,pc, in our sample (F606W filter, overplotted with ellipse
indicating $r_{h}$, derived via elliptical King model fit). Note
different image scales in panels (a) and (b). North is up, east is
left.}
\end{figure}

KW072 {\it (VdB0)} is located at the position of [15\arcmin,7\arcmin]
in Figure~7. A visual impression that it is surrounded by an enhanced
surface number density of stars (ejected from cluster) is confirmed by
the structural model fit, indicating its extended nature---structural
model parameters: ratio $r_{t}/r_{c} \sim 300$ (King), $n \sim 1$
(EFF), and $r_{h} \sim 5.5$\,pc. Its young age of $\sim$20\,Myr is
confirmed by a strong ultraviolet ({\it GALEX}) flux. A mass, $\log(m/m_{\sun})
\sim 4.6$, of the cluster, which formed recently, is remarkably
high comparing with the MW typical OCs' mass. This cluster was studied
in detail by \citet{Perina2009} based on photometry of individual
stars from $HST$ images. The parameters derived are in a very good
agreement with values derived in the present study based on its
integrated properties.

KW141 {\it (B011D)} is located at the position of [11\arcmin,15\arcmin]
in Figure~7. It represents a population of young, $\sim$10\,Myr,
intermediate-mass, $\log(m/m_{\sun}) \sim 3.8$, clusters. However,
stochastically appearing four bright red stars enclose the blueish
central part of this object suggesting an older age. Its $HST$ WFPC2
color image is displayed in Figure~8(a). The age and mass of this
``stochastic cluster example'' are in good agreement with
\citet{Caldwell2009}: $\sim$30\,Myr and $\log(m/m_{\sun}) \sim 3.8$.

KW249---the most controversial object, located close to the M31
bulge at the position of [3\arcmin,29\arcmin] in Figure~7. Its
young age of $\sim$10\,Myr and a mass of $\log(m/m_{\sun}) \sim 4.3$
make it incompatible with the structural parameter best-fit model
indicating very high ratio of $r_{t}/r_{c} > 500$ (King model) and
large half-light radius of $r_{h} \sim 14$\,pc (the most extended
object in our cluster sample). It is associated with 24\,$\mu$m
({\it Spitzer}) and HI emission, probably, still being an embedded
cluster with $E(B-V) \sim 1.5$, which could explain the absence of
$H\alpha$ and ultraviolet emission. It was imaged with the $HST$
WFPC2 (F606W filter), displayed in Figure~8(b) (fitted elliptical models
provide the ellipticity of $\sim$0.75---a ratio of minor to major
axis). We suggest, that it could be either: (1) an intermediate-mass
cluster, which expanded after a rapid gas removal phase \citep{Baumgardt2007},
or (2) a background galaxy, having emission in $K_{\rm s}$ (2MASS)
attributed to older stellar populations.

KW271 \& KW273 {\it (G099*)}---a double cluster candidate, located
at the position of [2\arcmin,18\arcmin] in Figure~7. KW271 is centered
on a strong 24\,$\mu$m emission source, its parameters are: $t \sim
10$\,Myr, $\log(m/m_{\sun}) \sim 3.7$, $E(B-V) \sim 0.9$, and $r_{h}
\sim 3.6$\,pc. Contrarily, KW273 has no 24\,$\mu$m emission, is much
older ($t \sim 800$\,Myr), more massive ($\log(m/m_{\sun}) \sim 4.5$),
has a low color excess ($E(B-V) \sim 0.1$), and large half-light radius
($r_{h} \sim 4.3$\,pc). The structural parameters of two ``adhered''
cluster candidates were derived by masking KW271 and fitting the
brighter KW273. After subtracting the {\it best-fit} model of KW273
from the original image, KW271 was fitted. In total, seven iterations
of the object subtraction were performed, leading to fine residual
images. We suggest two possible explanations of the nature of these
objects: (1) they make a double cluster, similar to those detected
in LMC/SMC \citep{Carvalho2008}, with one component just emerging
from the embedded phase. Close image inspection reveals an extended
``halo'' enshrouding both these objects, however, the derived age
difference does not support this case; (2) both clusters (they could
be a cluster and a galaxy---24\,$\mu$m emission source) are subject
to projection, however, probability of such close coincidence is
low. We note, that \citet{Caldwell2009} classify KW271 as a background
galaxy based on the measured redshift $z \sim 0.15$, and for KW273
they provide age of $\sim$400\,Myr and mass $\log(m/m_{\sun}) \sim
4.0$ in reasonable agreement with our values.

KW279 {\it (B205D)} is located at a large projected galactocentric
distance, thus far from the star-forming ring, at the position of
[2\arcmin,5\arcmin] in Figure~7. It is associated with a strong
24\,$\mu$m emission, which supports its young age of $\sim$10\,Myr.
A low-mass, $\log(m/m_{\sun}) \sim 3.2$, substantial color excess,
$E(B-V) \sim 0.6$, and a moderate size, $r_{h} \sim 2$\,pc, were
deduced assuming it being a star cluster. However, its asymmetric
shape with a blueish ``tail,'' pointing at a neighboring 24\,$\mu$m
emission source, implies that it could be an interacting starburst
galaxy. We note, that \citet{Caldwell2009} classify KW279 as a
background galaxy based on the measured redshift $z \sim 0.1$.

\LongTables
\begin{deluxetable}{c c c c c c c c c c}
\tabletypesize{\scriptsize}
\setlength{\tabcolsep}{4.0pt} \tablewidth{360pt} \tablecolumns{9}
\tablecaption{Evolutionary and Structural Parameters of {\it
High-Probability} Star Cluster Candidates in the Southwest Part of
the M31 Disk}
\tablehead{
\colhead{ID} &
\colhead{$\alpha_{\rm J2000}$} &
\colhead{$\delta_{\rm J2000}$} &
\colhead{$M_{V}$\tablenotemark{a}} &
\colhead{$\log(t/{\rm Myr})$\tablenotemark{b}} &
\colhead{$\log(m/m_{\sun})$\tablenotemark{c}} &
\colhead{[M/H]\tablenotemark{d}} &
\colhead{$E(B-V)$\tablenotemark{e}} &
\colhead{$r_{h}$\tablenotemark{f}}}
\startdata
KW003 &  10.04519 &  40.89759 &   -6.0 &   1.89 &   3.42 &   -0.5 &   0.12 &    2.0 \\
KW004 &  10.04568 &  40.60326 &   -6.6 &   1.67 &   3.48 &   -0.9 &   0.10 &    0.7 \\
KW006 &  10.05430 &  40.60535 &   -6.7 &   1.79 &   3.66 &   -0.2 &   0.21 &    0.7 \\
KW007 &  10.05468 &  40.69164 &   -6.1 &   2.04 &   3.56 &   -0.4 &   0.52 &    0.9 \\
KW008 &  10.05671 &  40.63810 &   -6.1 &   1.52 &   3.16 &   -0.8 &   0.30 &    0.6 \\
KW009 &  10.05700 &  40.76773 &   -5.2 &   1.36 &   2.79 &   -0.1 &   0.34 &    0.8 \\
KW010 &  10.05768 &  40.67549 &   -7.3 &   2.07 &   4.01 &   -0.5 &   0.70 &    5.1 \\
KW011 &  10.05924 &  40.65584 &   -6.6 &   1.88 &   3.66 &   -0.8 &   0.16 &    1.4 \\
KW013 &  10.06067 &  40.62244 &   -6.6 &   1.78 &   3.61 &   -0.4 &   0.02 &    1.5 \\
KW014 &  10.06177 &  40.77234 &   -6.3 &   1.42 &   3.26 &   -0.1 &   0.50 &    2.2 \\
KW015 &  10.06203 &  40.75771 &   -4.9 &   2.85 &   3.70 &    0.1 &   0.05 &    1.9 \\
KW017 &  10.06453 &  40.66658 &   -6.9 &   1.72 &   3.64 &   -0.7 &   0.22 &    1.0 \\
KW019 &  10.06504 &  40.58820 &   -5.2 &   2.56 &   3.54 &   -0.3 &   0.13 &    1.3 \\
KW020 &  10.06519 &  40.59857 &   -5.1 &   3.39 &   4.07 &   -1.1 &   0.09 &    2.6 \\
KW022 &  10.07194 &  40.65138 &   -7.0 &   1.74 &   3.72 &   -0.4 &   0.03 &    3.0 \\
KW023 &  10.07237 &  40.79158 &   -6.8 &   1.60 &   3.55 &   -0.5 &   0.41 &    2.7 \\
KW024 &  10.07249 &  40.54608 &   -5.1 &   1.59 &   2.92 &   -0.1 &   0.20 &    1.2 \\
KW025 &  10.07289 &  40.58077 &   -7.0 &   3.71 &   5.02 &   -1.9 &   0.52 &    8.7 \\
KW026 &  10.07313 &  40.65559 &   -6.6 &   1.83 &   3.63 &   -0.8 &   0.05 &    1.5 \\
KW027 &  10.07362 &  40.55231 &   -6.1 &   1.89 &   3.51 &    0.0 &   0.25 &    2.6 \\
KW028 &  10.07611 &  40.54573 &   -7.0 &   2.89 &   4.46 &   -0.7 &   0.04 &    8.1 \\
KW029 &  10.07649 &  40.66179 &   -5.0 &   1.61 &   2.85 &   -0.4 &   0.02 &    1.5 \\
KW031 &  10.07849 &  40.66804 &   -6.3 &   1.42 &   3.18 &   -0.7 &   0.04 &    0.8 \\
KW032 &  10.07944 &  40.63540 &   -6.6 &   1.62 &   3.51 &   -0.1 &   0.39 &    1.5 \\
KW033 &  10.08092 &  40.62481 &   -7.4 &   1.73 &   3.90 &   -0.4 &   0.30 &    1.7 \\
KW034 &  10.08174 &  40.58977 &   -5.7 &   1.74 &   3.25 &    0.1 &   0.32 &    2.0 \\
KW035 &  10.08174 &  40.71139 &   -6.1 &   1.91 &   3.46 &   -0.7 &   0.19 &    2.5 \\
KW037 &  10.08296 &  40.51322 &   -6.3 &   2.00 &   3.59 &   -0.5 &   0.07 &    0.7 \\
KW038 &  10.08386 &  40.50941 &   -5.5 &   2.85 &   3.81 &   -0.9 &   0.14 &    1.3 \\
KW039 &  10.08462 &  40.73291 &   -9.3 &   3.82 &   6.15 &   -0.6 &   0.15 &    1.2 \\
KW040 &  10.08659 &  40.55609 &   -6.5 &   1.97 &   3.66 &   -0.5 &   0.15 &    1.0 \\
KW041 &  10.08782 &  40.62758 &   -4.7 &   1.96 &   2.92 &   -0.4 &   0.03 &    1.3 \\
KW042 &  10.08929 &  40.62074 &   -5.9 &   2.69 &   3.92 &   -0.6 &   0.21 &    2.2 \\
KW043 &  10.09325 &  40.63796 &   -6.3 &   1.40 &   3.24 &   -0.2 &   0.35 &    0.7 \\
KW044 &  10.09623 &  40.51320 &   -9.2 &   1.77 &   4.59 &   -0.8 &   0.60 &   10.0 \\
KW045 &  10.09674 &  40.62105 &   -5.9 &   2.27 &   3.68 &   -0.1 &   0.22 &    1.1 \\
KW046 &  10.09793 &  40.64923 &   -6.3 &   1.45 &   3.19 &   -0.8 &   0.18 &    0.7 \\
KW048 &  10.10042 &  40.60629 &   -8.4 &   1.30 &   4.02 &   -0.2 &   0.19 &    0.8 \\
KW049 &  10.10088 &  40.53335 &   -6.0 &   1.92 &   3.44 &   -0.4 &   0.03 &    1.5 \\
KW050 &  10.10110 &  40.68546 &   -5.4 &   1.72 &   3.07 &   -0.4 &   0.03 &    1.7 \\
KW051 &  10.10121 &  40.61724 &   -4.5 &   2.26 &   3.13 &    0.1 &   0.02 &    0.9 \\
KW053 &  10.10341 &  40.81300 &   -7.6 &   2.40 &   4.40 &   -0.5 &   0.37 &    7.5 \\
KW054 &  10.10363 &  40.81692 &   -7.8 &   2.01 &   4.20 &   -0.4 &   0.69 &    4.6 \\
KW055 &  10.10418 &  40.55488 &   -5.5 &   1.91 &   3.24 &   -0.4 &   0.02 &    2.4 \\
KW056 &  10.10534 &  40.81352 &   -6.9 &   2.41 &   4.15 &   -0.4 &   0.38 &    2.8 \\
KW057 &  10.10801 &  40.62815 &   -9.8 &   0.84 &   3.97 &   -0.5 &   1.15 &    6.8 \\
KW058 &  10.10906 &  40.75886 &   -6.1 &   1.69 &   3.34 &   -0.2 &   0.47 &    1.0 \\
KW059 &  10.11086 &  40.79259 &   -7.4 &   2.11 &   4.05 &   -0.8 &   0.46 &    1.8 \\
KW061 &  10.11223 &  40.53262 &   -7.3 &   1.99 &   3.99 &   -0.4 &   0.63 &    2.6 \\
KW062 &  10.11367 &  40.75677 &   -6.7 &   1.98 &   3.73 &   -0.6 &   0.12 &    1.4 \\
KW064 &  10.11449 &  40.66775 &   -5.9 &   1.74 &   3.28 &   -0.5 &   0.24 &    1.1 \\
KW065 &  10.11469 &  40.51168 &   -4.9 &   2.46 &   3.37 &    0.0 &   0.03 &    3.2 \\
KW066 &  10.11525 &  40.52838 &   -5.6 &   1.48 &   2.99 &   -1.0 &   0.05 &    0.9 \\
KW067 &  10.11554 &  40.77210 &   -7.7 &   1.04 &   3.48 &   -0.1 &   0.93 &    1.3 \\
KW068 &  10.11716 &  40.54411 &   -5.0 &   2.71 &   3.61 &    0.0 &   0.04 &    3.1 \\
KW070 &  10.12135 &  40.85848 &   -6.3 &   3.77 &   5.08 &   -0.1 &   0.30 &    9.9 \\
KW071 &  10.12160 &  40.62482 &   -5.4 &   1.78 &   3.13 &   -0.8 &   0.15 &    3.9 \\
KW072 &  10.12255 &  40.60420 &  -10.0 &   1.23 &   4.63 &   -0.7 &   0.21 &    5.6 \\
KW073 &  10.12259 &  40.77968 &   -4.7 &   1.49 &   2.61 &   -0.5 &   0.04 &    0.9 \\
KW075 &  10.12328 &  40.54568 &   -5.2 &   3.12 &   3.84 &   -1.2 &   0.06 &    1.5 \\
KW076 &  10.12387 &  40.63842 &   -5.8 &   2.58 &   3.81 &   -0.4 &   0.30 &    1.7 \\
KW077 &  10.12448 &  40.51532 &   -5.7 &   2.55 &   3.74 &   -0.2 &   0.27 &    3.2 \\
KW078 &  10.12708 &  40.75818 &   -8.5 &   1.18 &   3.97 &   -0.5 &   0.03 &    1.0 \\
KW079 &  10.12764 &  40.74831 &   -7.1 &   1.11 &   3.54 &    0.5 &   0.03 &    1.7 \\
KW081 &  10.12861 &  40.83648 &   -6.7 &   2.40 &   4.04 &   -0.5 &   0.50 &    4.3 \\
KW082 &  10.12872 &  40.78416 &   -5.6 &   2.03 &   3.38 &   -0.3 &   0.07 &    0.8 \\
KW084 &  10.13069 &  40.75239 &   -5.1 &   0.95 &   2.52 &    0.4 &   0.03 &    1.2 \\
KW085 &  10.13216 &  40.51760 &   -4.8 &   2.30 &   3.21 &   -0.3 &   0.17 &    1.6 \\
KW086 &  10.13340 &  40.74484 &   -5.9 &   0.70 &   2.44 &   -0.0 &   0.09 &    0.9 \\
KW087 &  10.13404 &  40.77417 &   -6.0 &   3.76 &   4.64 &   -1.8 &   0.26 &    7.4 \\
KW089 &  10.13572 &  40.83714 &   -6.3 &   2.29 &   3.83 &   -0.3 &   0.07 &    2.2 \\
KW094 &  10.14379 &  40.80833 &   -6.7 &   1.11 &   3.19 &    0.1 &   0.65 &    0.7 \\
KW096 &  10.14847 &  40.93010 &   -6.8 &   1.18 &   3.30 &   -0.5 &   0.83 &    1.0 \\
KW097 &  10.14988 &  40.65192 &   -6.9 &   1.81 &   3.73 &   -0.9 &   0.47 &    1.1 \\
KW099 &  10.15130 &  40.84943 &   -6.1 &   3.44 &   4.52 &   -1.2 &   0.18 &    3.1 \\
KW100 &  10.15220 &  40.67090 &   -8.0 &   1.94 &   4.27 &   -0.4 &   0.51 &    5.6 \\
KW101 &  10.15475 &  40.55608 &   -6.2 &   2.82 &   4.08 &   -1.0 &   0.42 &    1.5 \\
KW102 &  10.15519 &  40.65407 &   -7.5 &   1.76 &   3.92 &   -1.0 &   0.55 &    1.3 \\
KW103 &  10.15562 &  40.81270 &   -6.4 &   2.23 &   3.84 &   -0.3 &   0.10 &    3.7 \\
KW105 &  10.15837 &  40.77468 &   -4.8 &   2.00 &   3.03 &   -0.3 &   0.06 &    1.5 \\
KW106 &  10.16050 &  40.74816 &   -5.5 &   3.11 &   3.96 &   -0.9 &   0.26 &    2.7 \\
KW107 &  10.16147 &  40.52894 &   -4.5 &   2.67 &   3.36 &   -0.1 &   0.08 &    1.1 \\
KW110 &  10.16421 &  40.80204 &   -4.9 &   3.02 &   3.74 &   -0.4 &   0.05 &    3.3 \\
KW112 &  10.17363 &  40.64115 &   -8.2 &   3.88 &   5.71 &   -1.2 &   0.51 &    6.0 \\
KW113 &  10.17612 &  40.60131 &   -6.5 &   3.38 &   4.67 &   -0.7 &   0.06 &    6.5 \\
KW114 &  10.17671 &  40.76094 &   -8.7 &   2.00 &   4.56 &   -0.3 &   1.16 &    2.9 \\
KW115 &  10.17808 &  40.83946 &   -5.3 &   2.05 &   3.27 &   -0.3 &   0.06 &    2.7 \\
KW119 &  10.18446 &  40.56793 &   -7.2 &   1.38 &   3.55 &   -0.6 &   0.48 &    2.8 \\
KW120 &  10.18520 &  40.62048 &   -8.8 &   1.94 &   4.57 &   -0.7 &   1.07 &    2.6 \\
KW121 &  10.18651 &  40.50836 &   -7.2 &   1.99 &   3.93 &   -0.6 &   0.65 &    3.4 \\
KW122 &  10.18702 &  40.88564 &   -6.5 &   2.58 &   4.06 &   -0.6 &   0.41 &    1.6 \\
KW123 &  10.19208 &  40.95922 &   -5.4 &   2.30 &   3.51 &    0.1 &   0.03 &    2.3 \\
KW124 &  10.19336 &  40.86136 &   -6.7 &   1.81 &   3.65 &   -0.8 &   0.30 &    2.5 \\
KW125 &  10.19582 &  40.68280 &   -5.7 &   3.42 &   4.38 &   -0.6 &   0.04 &    3.2 \\
KW126 &  10.19808 &  40.66878 &   -8.9 &   0.70 &   3.64 &    0.1 &   1.37 &    3.2 \\
KW127 &  10.19883 &  40.63281 &   -6.2 &   3.16 &   4.31 &   -0.4 &   0.24 &    5.9 \\
KW129 &  10.19919 &  40.92626 &   -5.8 &   2.38 &   3.71 &   -0.0 &   0.05 &    4.5 \\
KW130 &  10.20018 &  40.51726 &   -5.7 &   3.32 &   4.27 &   -0.8 &   0.07 &    2.9 \\
KW131 &  10.20031 &  40.67784 &   -6.9 &   2.31 &   4.04 &   -0.6 &   0.77 &    2.6 \\
KW132 &  10.20148 &  40.86623 &   -7.0 &   1.34 &   3.51 &   -0.1 &   0.23 &    1.3 \\
KW133 &  10.20148 &  40.58504 &   -6.8 &   2.31 &   3.99 &   -0.5 &   0.10 &    1.9 \\
KW134 &  10.20233 &  40.96013 &   -6.7 &   2.50 &   4.12 &   -0.2 &   0.33 &    2.1 \\
KW135 &  10.20362 &  40.56563 &   -6.5 &   1.83 &   3.64 &    0.0 &   0.19 &    6.5 \\
KW136 &  10.20492 &  40.64867 &   -5.5 &   2.93 &   3.90 &   -1.0 &   0.07 &    1.1 \\
KW137 &  10.20583 &  40.69227 &   -7.5 &   3.36 &   5.01 &   -1.0 &   0.18 &    4.2 \\
KW139 &  10.21178 &  40.67737 &   -6.0 &   3.20 &   4.36 &   -0.5 &   0.39 &    3.3 \\
KW140 &  10.21457 &  40.55768 &   -7.7 &   2.31 &   4.33 &   -0.6 &   0.47 &    6.7 \\
KW141 &  10.21504 &  40.73508 &   -8.7 &   0.84 &   3.81 &    0.6 &   0.65 &    0.8 \\
KW142 &  10.21700 &  40.58377 &   -5.8 &   1.85 &   3.30 &   -0.5 &   0.32 &    0.7 \\
KW143 &  10.21761 &  40.53484 &   -4.9 &   2.66 &   3.49 &   -0.5 &   0.06 &    3.1 \\
KW144 &  10.21773 &  40.97823 &   -6.6 &   2.78 &   4.21 &   -1.0 &   0.33 &    1.8 \\
KW145 &  10.21777 &  40.89900 &   -6.5 &   3.62 &   5.04 &   -0.1 &   0.07 &    8.8 \\
KW146 &  10.22036 &  40.86882 &   -6.2 &   3.20 &   4.33 &   -1.3 &   0.35 &    4.2 \\
KW147 &  10.22061 &  40.58877 &   -7.1 &   3.75 &   5.11 &   -1.9 &   0.34 &    5.8 \\
KW149 &  10.22197 &  40.73255 &   -6.5 &   0.70 &   2.66 &   -0.0 &   0.62 &    1.0 \\
KW150 &  10.22235 &  40.82742 &   -6.1 &   3.81 &   4.87 &   -0.6 &   0.42 &    3.8 \\
KW151 &  10.22301 &  40.71887 &   -5.9 &   2.37 &   3.70 &   -0.3 &   0.51 &    1.2 \\
KW152 &  10.22375 &  40.61417 &   -5.4 &   3.97 &   4.62 &   -1.4 &   0.24 &    1.1 \\
KW153 &  10.22448 &  40.67137 &   -4.6 &   2.75 &   3.49 &   -0.1 &   0.06 &    1.2 \\
KW154 &  10.22840 &  40.59210 &   -5.5 &   2.64 &   3.68 &   -0.8 &   0.14 &    1.6 \\
KW155 &  10.22846 &  40.73891 &   -5.4 &   3.37 &   4.19 &   -0.8 &   0.09 &    2.0 \\
KW156 &  10.23071 &  40.62244 &   -5.9 &   1.90 &   3.47 &    0.2 &   0.49 &    2.1 \\
KW158 &  10.23358 &  40.70630 &   -7.0 &   2.20 &   3.99 &   -0.6 &   0.62 &    1.6 \\
KW159 &  10.23475 &  40.57826 &   -5.3 &   2.34 &   3.42 &   -0.6 &   0.21 &    1.2 \\
KW160 &  10.23602 &  40.57370 &   -5.9 &   2.48 &   3.77 &   -0.6 &   0.06 &    3.6 \\
KW161 &  10.23726 &  40.60918 &   -7.3 &   2.45 &   4.29 &   -0.5 &   0.83 &    2.2 \\
KW164 &  10.23953 &  40.74093 &   -7.8 &   4.01 &   5.58 &   -1.9 &   0.59 &    1.8 \\
KW165 &  10.23976 &  40.54874 &   -6.2 &   1.58 &   3.32 &   -0.2 &   0.25 &    1.9 \\
KW166 &  10.24050 &  40.57272 &   -5.3 &   1.43 &   2.84 &   -1.0 &   0.06 &    0.7 \\
KW168 &  10.24321 &  40.89079 &   -9.3 &   0.70 &   3.81 &    0.4 &   1.49 &    2.3 \\
KW169 &  10.24352 &  40.60981 &   -6.9 &   1.50 &   3.58 &   -0.0 &   0.39 &    0.8 \\
KW170 &  10.24504 &  40.57345 &   -6.2 &   1.85 &   3.59 &    0.3 &   0.27 &    0.9 \\
KW171 &  10.24531 &  40.59666 &  -10.6 &   4.00 &   6.72 &   -1.6 &   0.14 &    3.7 \\
KW172 &  10.24540 &  40.71805 &   -6.4 &   3.83 &   5.23 &   -0.0 &   0.51 &    4.9 \\
KW173 &  10.24663 &  40.58450 &   -5.9 &   1.86 &   3.34 &   -0.7 &   0.26 &    2.4 \\
KW174 &  10.24728 &  40.56506 &   -5.3 &   2.38 &   3.48 &   -0.4 &   0.18 &    1.1 \\
KW176 &  10.25111 &  40.57447 &   -7.4 &   0.90 &   3.09 &   -0.3 &   0.81 &    1.4 \\
KW177 &  10.26062 &  40.84083 &   -5.0 &   2.12 &   3.14 &   -0.5 &   0.19 &    1.0 \\
KW178 &  10.26071 &  40.80432 &   -6.8 &   4.02 &   5.51 &   -0.0 &   0.34 &    3.2 \\
KW180 &  10.26142 &  40.56423 &   -6.1 &   2.03 &   3.56 &   -0.5 &   0.14 &    1.8 \\
KW181 &  10.26186 &  40.69644 &   -5.9 &   2.84 &   4.00 &   -0.9 &   0.36 &    1.9 \\
KW182 &  10.26203 &  40.58293 &   -6.5 &   1.93 &   3.65 &   -0.4 &   0.14 &    1.2 \\
KW183 &  10.26229 &  40.94328 &   -5.6 &   4.09 &   5.11 &   -0.0 &   0.08 &    9.7 \\
KW184 &  10.26513 &  40.92902 &   -6.3 &   3.30 &   4.45 &   -1.4 &   0.36 &    5.8 \\
KW185 &  10.26663 &  40.53928 &   -6.8 &   2.26 &   3.96 &   -0.5 &   0.43 &    9.2 \\
KW186 &  10.26836 &  40.66409 &   -5.2 &   2.98 &   3.84 &   -0.5 &   0.12 &    1.0 \\
KW187 &  10.26897 &  40.57899 &   -6.5 &   1.23 &   3.22 &   -0.1 &   0.20 &    1.2 \\
KW188 &  10.27108 &  40.75779 &   -6.1 &   2.46 &   3.85 &   -0.4 &   0.38 &    2.1 \\
KW189 &  10.27233 &  40.85868 &   -5.2 &   2.23 &   3.32 &   -0.5 &   0.07 &    1.5 \\
KW190 &  10.27249 &  40.58317 &   -6.8 &   1.54 &   3.60 &    0.1 &   0.21 &    0.9 \\
KW191 &  10.27438 &  40.89986 &   -6.0 &   2.65 &   3.90 &   -0.8 &   0.24 &    4.3 \\
KW192 &  10.27591 &  40.61567 &   -6.1 &   1.82 &   3.41 &   -0.8 &   0.35 &    3.7 \\
KW193 &  10.27601 &  40.60383 &   -5.3 &   2.39 &   3.51 &    0.1 &   0.04 &    3.6 \\
KW194 &  10.27857 &  40.57474 &   -6.9 &   1.80 &   3.70 &   -0.8 &   0.24 &    2.8 \\
KW195 &  10.28042 &  40.90625 &   -4.7 &   3.02 &   3.66 &   -0.4 &   0.04 &    1.8 \\
KW196 &  10.28306 &  40.88364 &   -6.9 &   4.01 &   5.32 &   -0.8 &   0.11 &    2.7 \\
KW197 &  10.28319 &  40.81240 &   -6.5 &   1.89 &   3.61 &   -0.7 &   0.45 &    5.4 \\
KW198 &  10.28443 &  40.54785 &   -7.5 &   2.10 &   4.16 &   -0.4 &   0.51 &    4.5 \\
KW199 &  10.28814 &  40.59810 &   -8.4 &   3.84 &   5.81 &   -0.7 &   0.09 &    2.2 \\
KW200 &  10.29161 &  40.96981 &   -7.0 &   3.19 &   4.66 &   -0.9 &   0.16 &    5.7 \\
KW201 &  10.29176 &  40.53308 &   -4.8 &   2.50 &   3.40 &   -0.0 &   0.04 &    3.3 \\
KW203 &  10.29775 &  40.80557 &   -6.7 &   1.34 &   3.44 &    0.2 &   0.29 &    1.6 \\
KW204 &  10.29872 &  40.58854 &   -5.3 &   2.32 &   3.42 &   -0.4 &   0.16 &    1.4 \\
KW205 &  10.29920 &  40.73882 &   -5.0 &   3.02 &   3.85 &    0.1 &   0.04 &    4.5 \\
KW206 &  10.29984 &  40.96378 &   -6.9 &   2.22 &   3.94 &   -0.6 &   0.62 &    4.0 \\
KW207 &  10.30204 &  40.67383 &   -6.9 &   2.09 &   3.88 &   -0.6 &   0.66 &    2.9 \\
KW208 &  10.30326 &  40.57155 &   -7.7 &   1.75 &   4.01 &   -0.3 &   0.24 &    2.9 \\
KW209 &  10.30461 &  40.64758 &   -4.9 &   2.76 &   3.60 &   -0.0 &   0.06 &    2.6 \\
KW210 &  10.30754 &  40.56614 &   -7.3 &   2.24 &   4.19 &   -0.4 &   0.06 &    9.3 \\
KW211 &  10.31052 &  40.93089 &   -9.3 &   4.00 &   6.19 &   -1.6 &   0.17 &    2.2 \\
KW212 &  10.31492 &  40.81869 &   -5.2 &   0.84 &   2.09 &   -0.2 &   0.09 &    1.3 \\
KW213 &  10.31781 &  40.86897 &   -5.0 &   2.93 &   3.69 &   -0.9 &   0.03 &    2.1 \\
KW214 &  10.31870 &  40.98434 &   -8.2 &   4.10 &   5.78 &   -2.2 &   0.19 &    1.7 \\
KW215 &  10.31888 &  40.65087 &   -6.5 &   1.84 &   3.57 &   -0.8 &   0.14 &    1.4 \\
KW216 &  10.31985 &  40.80730 &   -4.5 &   2.53 &   3.30 &   -0.0 &   0.05 &    2.5 \\
KW217 &  10.32373 &  40.72907 &   -5.8 &   2.49 &   3.75 &   -0.5 &   0.32 &    1.4 \\
KW218 &  10.32443 &  40.72654 &   -7.9 &   0.70 &   3.16 &   -0.3 &   1.09 &    2.9 \\
KW219 &  10.32559 &  40.73372 &   -6.7 &   2.79 &   4.28 &   -0.5 &   0.20 &    4.0 \\
KW221 &  10.32802 &  40.95444 &   -8.1 &   3.93 &   6.03 &    0.1 &   0.35 &    1.7 \\
KW223 &  10.32991 &  40.75117 &   -6.8 &   3.98 &   5.20 &   -1.4 &   0.26 &    1.3 \\
KW224 &  10.33554 &  40.60722 &   -4.3 &   1.84 &   2.70 &   -0.4 &   0.02 &    1.1 \\
KW225 &  10.33715 &  40.98462 &   -7.2 &   3.95 &   5.60 &   -0.1 &   0.17 &    1.7 \\
KW226 &  10.33754 &  40.73732 &   -6.0 &   2.30 &   3.71 &   -0.4 &   0.41 &    1.4 \\
KW227 &  10.33773 &  40.70366 &   -6.4 &   2.40 &   3.92 &   -0.4 &   0.39 &    2.0 \\
KW228 &  10.34021 &  40.68455 &   -6.2 &   2.23 &   3.72 &   -0.5 &   0.28 &    1.6 \\
KW229 &  10.34250 &  40.83295 &   -6.5 &   3.79 &   4.89 &   -1.9 &   0.48 &    4.8 \\
KW230 &  10.34662 &  40.73745 &   -4.4 &   2.88 &   3.47 &   -0.4 &   0.05 &    1.4 \\
KW231 &  10.34725 &  40.63251 &   -5.4 &   2.42 &   3.54 &   -0.1 &   0.24 &    3.4 \\
KW232 &  10.35035 &  40.61306 &   -6.8 &   1.98 &   3.79 &   -0.5 &   0.06 &    8.2 \\
KW233 &  10.35365 &  40.75562 &   -5.0 &   2.92 &   3.85 &    0.3 &   0.07 &    1.3 \\
KW234 &  10.35418 &  40.87940 &   -4.6 &   3.04 &   3.77 &    0.3 &   0.07 &    2.2 \\
KW235 &  10.35582 &  40.51488 &   -6.5 &   3.82 &   5.07 &   -0.5 &   0.56 &    2.3 \\
KW236 &  10.35815 &  40.56083 &   -5.5 &   2.58 &   3.69 &   -0.4 &   0.20 &    3.8 \\
KW239 &  10.36204 &  40.78528 &   -6.6 &   0.95 &   3.03 &    0.1 &   0.61 &    3.1 \\
KW240 &  10.36244 &  40.69372 &   -6.8 &   2.06 &   3.81 &   -0.4 &   0.31 &    1.8 \\
KW242 &  10.36512 &  40.80364 &   -6.3 &   2.02 &   3.60 &   -0.4 &   0.48 &    0.9 \\
KW243 &  10.36531 &  40.67627 &   -5.0 &   3.01 &   3.75 &   -0.5 &   0.04 &    1.1 \\
KW244 &  10.36715 &  40.89721 &   -9.1 &   3.86 &   6.22 &   -0.3 &   0.04 &    1.8 \\
KW245 &  10.37280 &  40.75469 &   -6.1 &   2.59 &   3.95 &   -0.3 &   0.14 &    4.2 \\
KW246 &  10.37317 &  40.83282 &   -7.8 &   1.88 &   4.19 &    0.1 &   0.94 &    3.2 \\
KW247 &  10.37492 &  40.64060 &   -5.7 &   1.76 &   3.25 &   -0.2 &   0.20 &    2.8 \\
KW248 &  10.37625 &  40.80908 &   -5.1 &   2.47 &   3.49 &    0.1 &   0.04 &    4.8 \\
KW249 &  10.37817 &  40.96697 &  -10.5 &   0.70 &   4.30 &    0.4 &   1.46 &   13.5 \\
KW250 &  10.37882 &  40.65230 &   -5.7 &   3.40 &   4.26 &   -1.6 &   0.24 &    2.9 \\
KW251 &  10.37981 &  40.77018 &   -5.7 &   2.56 &   3.79 &    0.1 &   0.03 &    2.2 \\
KW252 &  10.38217 &  40.67094 &   -6.5 &   1.49 &   3.34 &   -0.5 &   0.69 &    2.1 \\
KW253 &  10.38282 &  40.74846 &   -4.6 &   2.46 &   3.27 &    0.0 &   0.05 &    2.8 \\
KW254 &  10.38501 &  40.81791 &   -5.0 &   3.07 &   3.86 &    0.0 &   0.06 &    3.0 \\
KW255 &  10.38605 &  40.79066 &   -5.6 &   2.41 &   3.65 &   -0.1 &   0.32 &    2.9 \\
KW256 &  10.38768 &  40.65660 &   -5.6 &   1.89 &   3.25 &   -0.5 &   0.28 &    1.6 \\
KW257 &  10.38851 &  40.74583 &   -6.3 &   2.27 &   3.76 &   -0.5 &   0.55 &    3.9 \\
KW258 &  10.38910 &  40.92238 &   -9.9 &   0.70 &   4.05 &    0.4 &   1.66 &    3.4 \\
KW259 &  10.38952 &  40.66234 &   -4.7 &   2.57 &   3.39 &   -0.1 &   0.05 &    1.8 \\
KW260 &  10.38954 &  40.61910 &   -5.4 &   2.38 &   3.57 &    0.0 &   0.03 &    2.3 \\
KW261 &  10.38992 &  40.82127 &   -8.5 &   0.70 &   3.48 &   -0.0 &   1.39 &    2.1 \\
KW262 &  10.39174 &  40.63835 &   -5.2 &   2.73 &   3.72 &   -0.1 &   0.05 &    5.0 \\
KW263 &  10.39175 &  40.70641 &   -7.0 &   2.39 &   4.16 &   -0.4 &   0.68 &    2.8 \\
KW264 &  10.39327 &  40.87083 &   -5.8 &   2.39 &   3.73 &   -0.1 &   0.26 &    1.9 \\
KW265 &  10.39493 &  40.87377 &   -6.6 &   1.46 &   3.38 &   -0.2 &   0.59 &    1.1 \\
KW266 &  10.39585 &  40.76342 &   -6.0 &   3.17 &   4.24 &   -0.6 &   0.06 &    8.6 \\
KW267 &  10.39828 &  40.92745 &   -6.7 &   4.03 &   5.50 &    0.0 &   0.37 &    2.9 \\
KW268 &  10.39855 &  40.97107 &   -5.4 &   4.01 &   4.62 &   -1.9 &   0.16 &    3.5 \\
KW269 &  10.39923 &  40.58348 &   -7.4 &   1.08 &   3.43 &   -0.9 &   0.99 &    1.7 \\
KW270 &  10.40237 &  40.96336 &   -5.4 &   2.60 &   3.70 &    0.1 &   0.05 &    1.0 \\
KW271 &  10.40308 &  40.78978 &   -9.3 &   0.84 &   3.75 &   -0.2 &   0.92 &    5.4 \\
KW272 &  10.40315 &  40.84670 &   -9.0 &   1.73 &   4.50 &   -0.7 &   1.14 &    9.0 \\
KW273 &  10.40363 &  40.79043 &   -7.0 &   2.93 &   4.52 &   -0.5 &   0.06 &    6.1 \\
KW274 &  10.40468 &  40.94694 &   -4.9 &   3.97 &   4.73 &    0.1 &   0.07 &    3.5 \\
KW275 &  10.40531 &  40.68044 &   -5.0 &   2.76 &   3.67 &    0.1 &   0.04 &    2.7 \\
KW276 &  10.40555 &  40.88961 &   -5.0 &   2.41 &   3.38 &   -0.1 &   0.06 &    2.3 \\
KW277 &  10.40689 &  40.81064 &   -6.6 &   2.30 &   3.88 &   -0.7 &   0.71 &    2.3 \\
KW278 &  10.40818 &  40.79498 &   -6.4 &   2.48 &   4.02 &   -0.2 &   0.32 &    2.0 \\
KW279 &  10.40857 &  40.56961 &   -7.8 &   0.90 &   3.23 &   -0.4 &   0.55 &    1.9 \\
KW280 &  10.41025 &  40.97424 &   -6.0 &   3.01 &   4.25 &    0.1 &   0.15 &    3.8 \\
KW281 &  10.41050 &  40.82679 &   -5.7 &   2.77 &   3.98 &    0.2 &   0.04 &    1.5 \\
KW283 &  10.41167 &  40.79971 &   -6.3 &   3.81 &   5.05 &   -0.3 &   0.37 &    8.1 \\
KW284 &  10.41180 &  40.68181 &   -7.4 &   2.02 &   4.06 &   -0.5 &   0.14 &    1.2 \\
KW285 &  10.41559 &  40.94267 &   -6.5 &   3.92 &   5.26 &   -0.1 &   0.31 &    2.8 \\
\enddata
\tablecomments{ID number, R.A. \& Decl. (J2000) coordinates of the
photometric aperture center in the USNO-B1.0 catalog system (degrees)
according to \citet{Narbutis2008}.}
\tablenotetext{a}{Absolute magnitude, $M_{V}$, corrected for aperture
correction.}
\tablenotetext{b}{Age, $t$ in Myr, $\log(t/{\rm Myr})$.}
\tablenotetext{c}{Mass, $m$ in solar mass units, $\log(m/m_{\sun})$.}
\tablenotetext{d}{Metallicity, [M/H].}
\tablenotetext{e}{Interstellar extinction in MW plus M31; color
excess, $E(B-V)$.}
\tablenotetext{f}{Average half-light radius in pc, $r_{h} = (r_{h}^{\rm
K} + r_{h}^{\rm E})/2$, derived from the King and EFF models.}
\end{deluxetable}

\end{document}